\documentclass[notitlepage,12pt]{article}

\usepackage[paper=letterpaper,hmargin=2.5cm,top=2.3cm,bottom=2.3cm,bindingoffset=0cm,twoside,hoffset=0pt,marginparsep=0pt,marginparwidth=0pt]{geometry}

\usepackage{placeins} 
\usepackage{nicematrix}
\usepackage{amssymb}
\usepackage{amsmath}
\usepackage{amsfonts}
\usepackage{graphicx}
\usepackage{xcolor}
\usepackage{colortbl}
\usepackage{xspace}
\usepackage{ulem}
\usepackage{wasysym}
\usepackage[toc,page]{appendix}
\usepackage{cancel}
\usepackage{slashed}
\usepackage{anyfontsize}
\usepackage[11pt]{moresize}

\usepackage{cite}

\usepackage{multirow,rotating}

\def\gsim{\raise0.3ex\hbox{$\;>$\kern-0.75em\raise-1.1ex\hbox{$\sim\;$}}}
\def\lsim{\raise0.3ex\hbox{$\;<$\kern-0.75em\raise-1.1ex\hbox{$\sim\;$}}}

\newcommand{\ba}[1]{\begin{eqnarray} \label{(#1)}}
\newcommand{\ea}{\end{eqnarray}}

\newcommand{\newb}[1]{{\cellcolor{gray!35}#1}}

\newcommand{\impb}[1]{{\cellcolor{gray!14}#1}}

\def\gsim{\raise0.3ex\hbox{$\;>$\kern-0.75em\raise-1.1ex\hbox{$\sim\;$}}}
\def\lsim{\raise0.3ex\hbox{$\;<$\kern-0.75em\raise-1.1ex\hbox{$\sim\;$}}}

\title{QCD Running in Lepton Number Violating\\ Meson and Tau Decays}

\author{Marcela Gonz\'alez$^1$\footnote{marcela.gonzalezpi@uv.cl},
Nicol\'as A. Neill$^2$\footnote{naneill@outlook.com}
\\[2.5ex]
\small
$^1$\textit{Instituto de F\'isica y Astronom\'ia, Universidad de Valpara\'iso,}\\
\small
\textit{Avenida Gran Breta\~na 1111, Valpara\'iso, Chile}\\
\small
$^2$\textit{Universidad de Tarapac\'a, Departamento de Ingenier\'ia El\'ectrica-Electr\'onica,}\\
\small
\textit{Arica, Chile}
}
\date{}

\begin{document}

\maketitle

\begin{abstract}
Below the electroweak scale, new physics that violates lepton number in two units ($\Delta L = 2$) and is mediated by heavy particle exchange can be parameterized by a dimension-9 low-energy effective Lagrangian.
Operators in this Lagrangian involving first-generation quarks and leptons contribute to the short-range mechanism of neutrinoless double beta decay ($0\nu\beta\beta$) and therefore they are strongly constrained.
On the other hand, operators with other quark and lepton families are bounded by the non-observation of different lepton number violating (LNV) meson and tau decays, such as $M_1^- \to M_2^+\ell_1^-\ell_2^-$ and $\tau^- \to \ell^+ M_1^- M_2^-$.
In this work, we calculate RGE-improved bounds on the Wilson coefficients involved in these decays.
We calculate QCD corrections to the dimension-9 operator basis and find RG evolution matrices that describe the evolution of the Wilson coefficients across different energy scales. Unlike the running of operators involved in $0\nu\beta\beta$-decay, the general flavor structure leads to the mixing of not only different Lorentz structures but also of different quark-flavor configurations. Additionally, operators that vanish for the identical lepton case need to be added to the operator basis.
We find new constraints on previously unbounded operators and the enhancement of bounds for specific Wilson coefficients.
We also find new bounds coming from the mixing between operators with different quark-flavor configurations.
\end{abstract}

\newpage
\section{Introduction}
\label{sec:introduction}

The origin of neutrino masses and mixing remains an open question that requires physics beyond the standard model.
If neutrinos are Dirac fermions, the physics responsible for neutrino masses conserves total lepton number ($L$).
On the other hand, it is known that the existence of new physics that violates lepton number in two units ($\Delta L =2$) would imply the generation of Majorana mass terms at the loop level \cite{Schechter:1981bd}.
Therefore, the observation of a lepton number violating (LNV) process would shed light on the type of new physics responsible for neutrino masses and on the Dirac vs. Majorana nature of neutrinos.

Many experiments have searched for LNV in different processes.
The golden channel for the observation of LNV is neutrinoless double beta ($0\nu\beta\beta$) decay  \cite{Rodejohann:2011mu,Deppisch:2012nb}, which puts stringent bounds on the $0\nu\beta\beta$-decay half-life ($T_{1/2} \gtrsim 2.3\times 10^{26}\,\mbox{yr}$ \cite{KamLAND-Zen:2022tow}).
Other processes that are sensitive to lepton-number violation include meson decays such as $K^-\to \pi^+ \mu^- \mu^-$ \cite{NA62:2022tte, NA62:2019eax, NA62:2021zxl, LHCb:2020car, BaBar:2011ouc, BaBar:2012eip, LHCb:2014osd, BaBar:2013swg, LHCb:2011yaj, BELLE:2011bej, LHCb:2012pcm,
Littenberg:1991ek, Quintero:2011yh, Dong:2013raa, Liu:2016oph, Abada:2017jjx, Cvetic:2017vwl, Mandal:2017tab, Mejia-Guisao:2017gqp, Yuan:2017uyq, Chun:2019nwi, Godbole:2020jqw,
Deppisch:2020oyx}, tau-lepton decays such as $\tau^- \to e^+ \pi^- \pi^-$ \cite{Belle:2012unr,Belle:2009myz,Ilakovac:1995wc,LopezCastro:2012udb,Yuan:2017xdp}, $\ell_1^-\to \ell_2^+$ lepton conversion \cite{SINDRUMII:1998mwd,Simkovic:2000ma,Simkovic:2001fs,Berryman:2016slh,Yeo:2017fej}, and collider searches with same-sign leptons in the final state \cite{Allanach:2009xx,Helo:2013ika,Helo:2013dla,Peng:2015haa,Gonzalez:2016ztm}.
Unlike $0\nu\beta\beta$-decay, both meson and tau decays allow to prove LNV interactions that are not restricted to first generation quarks and leptons.

The new LNV physics can be described in a model-independent way by using low energy effective Lagrangians.
In particular, dimension-9 operators involving first-generation quarks and leptons have been used to parameterize the different contributions to $0\nu\beta\beta$-decay mediated by heavy particle exchange (the so-called short-range mechanism) \cite{Pas:2000vn}, allowing to find model-independent bounds on the corresponding Wilson coefficients.
Meson and tau decays have also been used to put tree-level constrains on dimension-9 LNV operators involving quarks and leptons not restricted to the first generation \cite{Quintero:2016iwi}. 
The effects of renormalization group evolution (RGE) have proven to be relevant when extracting bounds on the Wilson coefficients contributing to $0\nu\beta\beta$-decay \cite{Mahajan:2013ixa,Peng:2015haa,Gonzalez:2015ady,Arbelaez:2016uto,Arbelaez:2016zlt,Gonzalez:2017mcg,Ayala:2020gtv},
as well as putting bounds on new physics in other contexts, such as flavor physics \cite{Buras:1998raa},
meson and tau decays \cite{Liao:2019gex,Liao:2020roy,Gonzalez:2021tqc,Liao:2021qfj}, 
and dark matter direct detection \cite{DEramo:2016gos,Bishara:2018vix}.

In this work, we derive the renormalization group evolution for the complete dimension-9 LNV ($\Delta L =2$) operator basis contributing to the LNV meson and tau decays, $M_1^-\to M_2^+ \ell_1^- \ell_2^-$ and $\tau^-\to \ell^+ M_1^- M_2^-$, respectively.\footnote{The conjugate processes are implied here and throughout the text.}
This basis, with operators of the  form $\mathcal O\sim (\bar u_i\Gamma_1 d_j)(\bar u_k\Gamma_2 d_l)(\bar \ell_1 \Gamma_3 \ell_2^c)$, is a generalization of the basis considered in the short-range mechanism of $0\nu\beta\beta$-decay since it is not constrained to first-generation quarks, and in general $\ell_1 \neq \ell_2$.
As we will show in Sec.~(\ref{sec:QCDcorrections}), this introduces additional operators (not present for equal lepton flavors) and additional quark mixings.
We apply this results to extract RGE-improved bounds on the Wilson coefficients relevant to the meson and tau decays listed in Tables\,(\ref{tab:expbounds_mesondecays}) and (\ref{tab:expbounds_taudecays}).\footnote{{Present and upcoming experiments are expected improve these bounds \cite{Chun:2019nwi}. For instance, the High Intensity Kaon Experiments (HIKE) 
\cite{HIKE:2023ext} are expected to probe the branching fraction of $K^-\to \pi^+ \ell^- \ell^-$ with a sensitivity of $\mathcal O(10^{-13})$, while current constraints on $D$- and $B$-meson decays from LHCb, based on integrated luminosities of $\mathcal O(1\,\text{fb}^{-1})$, could improve by approximately two orders of magnitude as the target luminosity ($300\,\text{fb}^{-1}$) is achieved.
Similarly, current bounds derived from BaBar and Belle experiments could be improved by up to two orders of magnitude once the target luminosity of Belle-II ($50\,\text{ab}^{-1}$) is reached.
}
}

Our paper in organized as follows.
In Section\,(\ref{sec:LNVLag}) we present the dimension-9 low energy effective Lagrangian and review how tree-level bounds on Wilson coefficients are derived from LNV meson and tau decays.
In Section\,(\ref{sec:QCDcorrections}) we calculate QCD corrections to the dimension-9 LNV basis 
and find the anomalous dimension and evolution matrices.
These results are applied in Section\,(\ref{sec:RGEinducedbounds}) to find new RGE-induced bounds on the Wilson coefficients, which are the main result of our work together with the anomalous dimensions for the general case with arbitrary quark and lepton flavors.
Finally, our summary and conclusions are presented in Section\,(\ref{sec:conclusions}).

\begin{table}[b]
\centering
\scriptsize
\begin{tabular}{c|c||c|c||c|c}
Process  &  Limit on BR & Process  &  Limit on BR & Process  &  Limit on BR\\
\hline\hline
$K^- \to \pi^+ e^- e^-$ & $5.3\times 10^{-11}$ \cite{NA62:2022tte} &
$K^- \to \pi^+ \mu^- \mu^-$ & $4.2\times 10^{-11}$ \cite{NA62:2019eax} &
$K^- \to \pi^+ e^- \mu^-$ & $4.2\times 10^{-11}$ \cite{NA62:2021zxl}
\\
$D^- \to \pi^+ e^- e^-$ & $5.3\times 10^{-7}$ \cite{LHCb:2020car} &
$D^- \to \pi^+ \mu^- \mu^-$ & $1.4\times 10^{-8}$ \cite{LHCb:2020car} &
$D^- \to \pi^+ e^- \mu^-$ & $1.3\times 10^{-7}$ \cite{LHCb:2020car}
\\
$D^- \to K^+ e^- e^-$ & $9.0\times 10^{-7}$ \cite{BaBar:2011ouc} &
$D^- \to K^+ \mu^- \mu^-$ & $1.0\times 10^{-5}$ \cite{BaBar:2011ouc} &
$D^- \to K^+ e^- \mu^-$ & $1.9\times 10^{-6}$ \cite{BaBar:2011ouc}
\\
$D_s^- \to \pi^+ e^- e^-$ & $1.4\times 10^{-6}$ \cite{LHCb:2020car} &
$D_s^- \to \pi^+ \mu^- \mu^-$ & $8.6\times 10^{-8}$ \cite{LHCb:2020car} &
$D_s^- \to \pi^+ e^- \mu^-$ & $6.3\times 10^{-7}$ \cite{LHCb:2020car}
\\
$D_s^- \to K^+ e^- e^-$ & $7.7\times 10^{-7}$ \cite{LHCb:2020car} &
$D_s^- \to K^+ \mu^- \mu^-$ & $2.6\times 10^{-8}$ \cite{LHCb:2020car} &
$D_s^- \to K^+ e^- \mu^-$ & $2.6\times 10^{-7}$ \cite{LHCb:2020car}
\\
$B^- \to \pi^+ e^- e^-$ & $2.3\times 10^{-8}$ \cite{BaBar:2012eip} &
$B^- \to \pi^+ \mu^- \mu^-$ & $4.0\times 10^{-9}$ \cite{LHCb:2014osd} &
$B^- \to \pi^+ e^- \mu^-$ & $1.5\times 10^{-7}$ \cite{BaBar:2013swg}
\\
$B^- \to K^+ e^- e^-$ & $3.0\times 10^{-8}$ \cite{BaBar:2012eip} &
$B^- \to K^+ \mu^- \mu^-$ & $4.1\times 10^{-8}$ \cite{LHCb:2011yaj}&
$B^- \to K^+ e^- \mu^-$ & $1.6\times 10^{-7}$ \cite{BaBar:2013swg}  
\\
$B^- \to D^+ e^- e^-$ & $2.6\times 10^{-6}$ \cite{BaBar:2013swg,BELLE:2011bej}&
$B^- \to D^+ \mu^- \mu^-$ & $6.9\times 10^{-7}$ \cite{LHCb:2012pcm} &
$B^- \to D^+ e^- \mu^-$ & $1.8\times 10^{-6}$ \cite{BELLE:2011bej}
\\
 & &
$B^- \to D_s^+ \mu^- \mu^-$ & $5.8\times 10^{-7}$ \cite{LHCb:2012pcm} &
& \\
\hline
\end{tabular}
\caption{Current experimental upper bounds on the branching ratios (BR) of different LNV meson decays.
}\label{tab:expbounds_mesondecays}
\normalsize
\end{table}

\begin{table}[tb]
\centering
\small
\begin{tabular}{c|c||c|c}
Process  &  Limit on BR \cite{Belle:2012unr} & Process  &  Limit on BR \cite{Belle:2012unr}\\
\hline\hline
$\tau^-\to e^+ \pi^-\pi^-$ & $2.0\times 10^{-8}$ &
$\tau^-\to \mu^+ \pi^-\pi^-$ & $3.9\times 10^{-8}$ 
\\
$\tau^-\to e^+ \pi^-K^-$ & $3.2\times 10^{-8}$ &
$\tau^-\to \mu^+ \pi^-K^-$ & $4.8\times 10^{-8}$
\\
$\tau^-\to e^+ K^-K^-$ & $3.3\times 10^{-8}$ &
$\tau^-\to \mu^+ K^-K^-$ & $4.7\times 10^{-8}$
\\
\hline
\end{tabular}
\caption{Current experimental upper bounds on the branching ratios (BR) of different LNV tau decays.
}\label{tab:expbounds_taudecays}
\normalsize
\end{table}

\section{LNV Effective Lagrangian and LNV Decays}\label{sec:LNVLag}

The most general low energy dimension-9 effective Lagrangian that involves quark currents and violates lepton number in two units ($\Delta L =2$) can be written as
\begin{eqnarray}\label{eq:LagGen}
{\cal L}^{\Delta L = 2}_{\rm eff} = \frac{G_F^2}{2 m} \,
              \sum_{\substack{n, XYZ\\ijkl,\ell_1\ell_2}} C_{n(ijkl)}^{XYZ(\ell_1\ell_2)}(\mu)\cdot \mathcal{O}^{XYZ(\ell_1\ell_2)}_{n(ijkl)}(\mu),\label{eq:effL}
\end{eqnarray}
where $m$ is a mass scale that we will choose depending of the process in consideration, $C_{n(ijkl)}^{XYZ(\ell_1\ell_2)}$ are Wilson coefficients, and $\mathcal O_{n(ijkl)}^{XYZ(\ell_1\ell_2)}$ are a complete basis for the LNV ($\Delta L = 2$) dimension-9 operators:
\begin{eqnarray}
\mathcal{O}^{XYZ(\ell_1\ell_2)}_{1(ijkl)}&=& 8 ({\bar u_i}P_{X}d_j) ({\bar u_k}P_{Y}d_l) \ j^Z, \label{eq:OperBasis-1}\\
\mathcal{O}^{XXX(\ell_1\ell_2)}_{2(ijkl)}&=& 8 ({\bar u_i}\sigma^{\mu\nu}P_{X}d_j)
({\bar u_k}\sigma_{\mu\nu}P_{X}d_l) \ j^X, \label{eq:OperBasis-2}\\
\mathcal{O}^{XYZ(\ell_1\ell_2)}_{3(ijkl)}&=& 8 ({\bar u_i}\gamma^{\mu}P_{X}d_j)
({\bar u_k}\gamma_{\mu}P_{Y}d_l) \  j^Z,\label{eq:OperBasis-3}\\
\mathcal{O}^{XYZ(\ell_1\ell_2)}_{4(ijkl)}&=& 8 ({\bar u_i}\gamma^{\mu}P_{X}d_j) ({\bar u_k}\sigma_{\mu}^{\,\,\nu}P_{Y}d_l) \ j_{\nu}^Z,\label{eq:OperBasis-4}\\
\mathcal{O}^{XYZ(\ell_1\ell_2)}_{5(ijkl)}&=& 8 ({\bar u_i}\gamma^{\mu}P_{X}d_j) ({\bar u_k}P_{Y}d_l) \ j_{\mu}^Z.\label{eq:OperBasis-5}\\
\mathcal{O}^{XYZ(\ell_1\ell_2)}_{6(ijkl)}&=&
8 ({\bar u_i}\gamma^\mu P_{X}d_j)
({\bar u_k}\gamma^{\nu}P_{Y}d_l) \ j^Z_{\mu\nu}, \label{eq:OperBasis-6}\\
\mathcal{O}^{XZZ(\ell_1\ell_2)}_{7(ijkl)}&=&
8 ({\bar u_i}P_{X}d_j)
({\bar u_k}\sigma^{\mu\nu}P_{Z}d_l) \ j^Z_{\mu\nu}, \label{eq:OperBasis-7}\\
\mathcal{O}^{XXX(\ell_1\ell_2)}_{8(ijkl)}&=&
8 ({\bar u_i}\sigma^{\mu\alpha} P_{X}d_j)
({\bar u_k}\sigma^{\nu}_{\,\alpha}P_{X}d_l) \ j^X_{\mu\nu}, \label{eq:OperBasis-8}
\end{eqnarray}
with
\begin{eqnarray}
j^Z = \bar \ell_1 P_Z \ell^c_2,\ \ \ 
j^Z_\mu = \bar \ell_1 \gamma^\mu P_Z \ell^c_2,\ \ \
j_{\mu\nu}^Z = \bar \ell_1 \sigma^{\mu\nu} P_Z \ell^c_2,
\end{eqnarray}
where $P_X$ are chiral projectors with the indices $X,Y,Z=L,R$ the corresponding chiralities, while the indices $i,j=u,c$ and $k,l=d,s,b$ run over the quark flavors and  $\ell_1,\ell_2=e,\mu,\tau$ over lepton flavors.
The factor eight in front of the definition of the operators has been introduced for consistency with the basis without chiral projectors \cite{Pas:2000vn}.
Operators $\mathcal O_2$, $\mathcal O_7$, and $\mathcal O_8$ appear with repeated chiralities since $\sigma^{\mu\nu} P_X\otimes \sigma_{\mu\nu} P_Y = 0$ for $X\neq Y$ and $\sigma^{\mu\alpha} P_X \otimes \sigma^{\nu}_{\,\alpha} P_Y \otimes \sigma_{\mu\nu} P_Z$ is non vanishing only for $X=Y=Z$, as can be shown by Fierz transformations.
Additionally, for $\ell_1=\ell_2$, the operators $\mathcal O_{6,7,8}$ vanish since $\bar \ell \sigma^{\mu\nu} P_X \ell^c  \equiv 0$.
In this work, we consider meson decays with both same and different flavor leptons in the final state, and tau lepton decays that, due to kinematics, necessarily involve different lepton flavors.
Therefore, we will consider the full basis ($\mathcal O_n$, with $n=1,\ldots ,8$) shown in Eqs.\,(\ref{eq:OperBasis-1}-\ref{eq:OperBasis-8}).
For $\ell_1=\ell_2 = e$, the subset of operators $\mathcal O_{1,2,3,4,5}$ involving first generation quarks (i.e., $ijkl=udud$) contribute to the short-range mechanism of $0\nu\beta\beta$-decay, which has been already studied in the literature \cite{Pas:2000vn,Bonnet:2012kh,Gonzalez:2015ady}.
Given the strong bounds on the $0\nu\beta\beta$-decay lifetime, the Wilson coefficients $\mathcal O_{1,2,3,4,5(udud)}^{XYZ(ee)}$ are strongly constrained (with bounds of the order $\sim \mathcal O(10^{-7}) \to \mathcal O(10^{-9})$ depending on the operator  \cite{Pas:2000vn,Gonzalez:2015ady}).
On the other hand, Wilson coefficients with other quark flavors ($ijkl\neq udud$) or other lepton flavors $(\ell_1 \ell_2\neq ee)$ are not constrained by $0\nu\beta\beta$-decay.
These Wilson coefficients are constrained by the meson and tau decays considered in this work:
the meson decays $M_1^-\to M_2^+\ell_1^{-}\ell_2^{-}$ put bounds on operators with $ijkl\neq udud$ since
they necessarily include at least one non-first-generation quark, while the tau decays $\tau^-\to \ell^+ M_1^- M_2^-$ constrain operators involving different lepton flavors ($\ell_1\neq \ell_2$).

The tree-level amplitudes for the process $M_1^-\to M_2^+ \ell_1^- \ell_2^-$ mediated by the effective Lagrangian in Eq.\,(\ref{eq:effL}) are given by
\begin{align}
    \mathcal M(M_1^-\to M_2^+ \ell_1^{-} \ell_2^{-}) = \left<M_2^+ \ell_1^{-} \ell_2^{-}|\mathcal L_{eff}^{\Delta L=2}|M_1^- \right>
    =
    \frac{G_F^2}{2 m_{M_1}}\sum_{\substack{n, XYZ\\ijkl,\ell_1\ell_2}} C_{n(ijkl)}^{XYZ(\ell_1\ell_2)} \mathcal A_{n(ijkl)}^{XYZ(\ell_1\ell_2)},
\end{align}
where $\mathcal A_{n(ijkl)}^{XYZ(\ell_1\ell_2)} = \left<M_2^+ \ell_1^{-} \ell_2^{-}|\mathcal O_{n(ijkl)}^{XYZ(\ell_1\ell_2)}|M_1^- \right>$ and $M_1^-=(\bar u_i d_j)$, $M_2^+ =(u_k \bar d_l)$.
Parameterizing the hadronic matrix elements as
\begin{align}
    \left<0| \bar u_i \gamma^5 d_j |M \right>
    = i \xi_M f_M,\ \ \ \xi_M=\frac{m_M^2}{m_{u_i}+m_{d_j}}, \ \ \
    \left<0| \bar u_i \gamma^\mu \gamma^5 d_j |M \right>
    = i p^\mu f_M,\label{eq:hadronicME}
\end{align}
where $p^\mu$ is the momentum of the meson $M$, we find:
\begin{align}
    \mathcal  A_1^{XYZ} & = \pm 2 f_{M_1} f_{M_2} \xi_{M_1} \xi_{M_2} [\bar u(p_{\ell_1}) P_Z\, v(p_{\ell_2})],\label{eq:A1}\\
    \mathcal  A_3^{XYZ} & = \pm 2 f_{M_1} f_{M_2} (p_M\cdot p_M') [\bar u(p_{\ell_1}) P_Z\, v(p_{\ell_2})],\\
    \mathcal A_5^{XYZ} & = \pm 2 f_{M_1} f_{M_2} \xi_{M_2} p_M^\mu [\bar u(p_{\ell_1}) \gamma_\mu P_Z\, v(p_{\ell_2})],\\
    \mathcal A_6^{XYZ} & = \pm 2 f_{M_1} f_{M_2} p_{M_1}^\mu p_{M_2}^\nu [\bar u(p_{\ell_1}) \sigma_{\mu\nu} P_Z\, v(p_{\ell_2})]\label{eq:A6},
\end{align}
where in the previous expressions the global $+$ and $-$ signs correspond to the cases $X=Y$ and $X\neq Y$, respectively.
The amplitudes $\mathcal A_2$, $\mathcal A_4$. $\mathcal A_7$, and $\mathcal A_8$ are not present since meson decay modes mediated by tensor currents are 
suppressed (vanish to first order) \cite{deGouvea:2007qla,Quintero:2016iwi}.

The partial decay widths for the meson decays are given by \cite{Quintero:2016iwi}
\begin{align}
    \Gamma(M_1^-\to M_2^+ \ell_1^- \ell_2^-) = &
    \left(1-\frac{1}{2}\delta_{\ell_1\ell_2}\right) \frac{G_F^4}{128(2\pi)^3 m_{M_1}^5}\times \nonumber\\
    &
    \sum_{\substack{n, XYZ}} \left|C_{n(ijkl)}^{XYZ(\ell_1\ell_2)}\right| \int_{s^-}^{s^+} ds \int_{t^-}^{t^+} dt \left|\mathcal{\overline{A}}_{n(ijkl)}^{XYZ(\ell_1\ell_2)}\right|^2,\label{eq:mesondecaywidth}
\end{align}
where $\mathcal{\overline{A}}_n$ are the spin-averaged amplitudes found in Eqs.\,(\ref{eq:A1})-(\ref{eq:A6}), $s\equiv (p_{\ell_1}+p_{\ell_2})^2$, $t\equiv (p_{\ell_2}+p_{M_2})^2$, the integration limits are given by
\begin{align}
    s^{-} & = (m_{\ell_1}+m_{\ell_2})^2,\ \ \ s^{+} = (m_{M_1}-m_{M_2})^2,\\
    t^{\pm} & = m_{M_1}^2 + m_{\ell_1}^2 - \frac{1}{2s}\left[
    \left(s+m_{M_1}^2-m_{M_2}^2\right)
    \left(s+m_{M_{\ell_1}}^2-m_{M_{\ell_2}}^2\right)\right.\nonumber\\
    &
    \left.
    \ \ \ \ \ \
    \mp \lambda^{1/2}(s,m_{\ell_1}^2,m_{\ell_2}^2)
    \lambda^{1/2}(s,m_{M_1}^2,m_{M_2}^2)
    \right],
\end{align}
with $\lambda(x,y,z)=x^2+y^2+z^2-2(xy+xz+yz)$, and the quark flavor indices ($i,j,k,l$) are fixed by the quark content of the mesons: $M_1^-=(\bar u_i d_j)$, $M_2^+=(u_k \bar d_l)$.

Equivalently, for tau lepton decays, the tree-level amplitude for the process $\tau^-\to \ell^+ M_1^- M_2^-$ is given by
\begin{align}
    \mathcal M(\tau^- \to \ell^+ M_1^- M_2^-) = \left<\ell^+ M_1^- M_2^-|\mathcal L_{eff}^{\Delta L=2}|\tau^- \right>
    =
    \frac{G_F^2}{2 m_{\tau}}\sum_{n,XYZ} C_{n(ijkl)}^{XYZ(\tau\ell)} \mathcal T_{n(ijkl)}^{XYZ(\tau\ell)},
\end{align}
where $\mathcal T_{n(ijkl)}^{XYZ(\tau\ell)} = \left<\ell^+ M_1^- M_2^-|\mathcal O_{n(ijkl)}^{XYZ(\tau\ell)}|\tau^-\right>$ and $M_1^-=(\bar u_i d_j)$, $M_2^- =(\bar u_k d_l)$.
Using the same parameterization for the hadronic matrix elements [Eq.\,(\ref{eq:hadronicME})], we find similar expressions for $\mathcal T_n$ as for $\mathcal A_n$ in Eqs.\,(\ref{eq:A1})-(\ref{eq:A6}), with the replacement $p_{\ell_1} \to p_\tau$ and $p_{\ell_2} \to p_\ell$.
The partial decay widths for the tau lepton are given by \cite{Quintero:2016iwi}
\begin{align}
    \Gamma(\tau^- \to \ell^+ M_1^- M_2^-) = & 
    \left(1-\frac{1}{2}\delta_{M_1 M_2}\right) \frac{G_F^4}{256(2\pi)^3 m_{\tau}^5}\times \nonumber\\
    &
    \sum_{n,XYZ} \left|C_{n(ijkl)}^{XYZ(\tau\ell)}\right| \int_{\tilde s^-}^{\tilde s^+} d\tilde s \int_{\tilde t^-}^{\tilde t^+} d\tilde t \left|\mathcal{\overline{T}}_{n(ijkl)}^{XYZ(\tau\ell)}\right|^2,\label{eq:taudecaywidth}
\end{align}
where $\mathcal{\overline{T}}_n$ are the spin-averaged amplitudes, $\tilde s\equiv (p_{M_1}+p_{M_2})^2$, $\tilde t\equiv (p_{\ell}+p_{M_2})^2$, the integration limits are given by
\begin{align}
    \tilde s^{-} & = (m_{M_1}+m_{M_2})^2,\ \ \ s^{+} = (m_{\tau}-m_{\ell})^2,\\
    \tilde t^{\pm} & = m_{\tau}^2 + m_{M_1}^2 - \frac{1}{2\tilde s}\left[
    \left(\tilde s+m_{\tau}^2-m_{\ell}^2\right)
    \left(\tilde s+m_{M_{M_1}}^2-m_{M_{M_2}}^2\right)\right.\nonumber\\
    & \left. \ \ \ \ \ \ \mp \lambda^{1/2}(\tilde s,m_{\tau}^2,m_{\ell}^2)
    \lambda^{1/2}(\tilde s,m_{M_1}^2,m_{M_2}^2)
    \right],
\end{align}
where the quark flavor indices ($i,j,k,l$) are fixed by the quark content of the final mesons: $M_1^-=(\bar u_i d_j)$, $M_2^-=(\bar u_k d_l)$.

Eqs.\,(\ref{eq:mesondecaywidth}) and (\ref{eq:taudecaywidth}) allow to find tree-level bounds on the Wilson coefficients $C_{n(ijkl)}^{XYZ(\ell_1\ell_2)}$, which we will use in Sec.\,(\ref{sec:RGEinducedbounds}) to find RGE-improved bounds.

\section{QCD running of WC for $|\Delta L|=2$ Processes}\label{sec:QCDcorrections}

In this section, we derive the QCD running of the WCs corresponding to the LNV ($\Delta L=2$) effective operator basis presented in Ecs.\,(\ref{eq:OperBasis-1})-(\ref{eq:OperBasis-8}), which contributes to the meson and tau decays discussed in the previous sections ($M_1^-\to M_2^{+} \ell_1^- \ell_2^-$ and $\tau^-\to \ell^+ M_1^-M_2^{-}$).

The formalism used to calculate QCD corrections in this case is analogous to the one developed in Ref. \cite{Gonzalez:2015ady} in the context of $0\nu\beta\beta$-decay, based in Refs. \cite{Buchalla:1995vs,Buras:1998raa}.
Here we briefly discuss the differences and additional aspects that need to be taken into account for the processes studied in this work.
The flavour configuration in the quark-level Lagrangian of $0\nu\beta\beta$-decay is composed by only two quark flavors ($u$ and $d$), having schematically operators of the form $(\bar{u}\Gamma_1 d)(\bar{u} \Gamma_2 d) \cdot j_{\ell}$, with $\Gamma_1,\Gamma_2$ representing in general different Lorentz structures, and $j_{\ell}$ the leptonic current, which is irrelevant from the QCD point of view in the case of $0\nu\beta\beta$-decay, but not in the general case where the chirality of the tensor lepton current does impact operator mixing, as we will see.
Fierz transformations and mixing under renormalization of these kind of operators lead back to the same flavor structure $(\bar{u}\Gamma_1 d)(\bar{u} \Gamma_2 d)$ but with the consequent mixing between different Lorentz structures.
This is not necessarily the case for the operators that contribute to $M_1^-\to M_2^{+} \ell_1^- \ell_2^-$ and $\tau^-\to \ell^+ M_1^-M_2^{-}$ decays.
Depending on the specific process under consideration, a generally non-invariant flavor structure under Fierz transformation can result in mixing not only among operators with different Lorentz structures but also with distinct flavor configurations.
Such behavior leads to anomalous dimension and $\mu$-evolution matrices that are, in general, of larger size than those in the $0\nu\beta\beta$-decay case.
From the perspective of meson decays, this implies that after hadronization, mixing will emerge between operators contributing to the decay of different mesons.
Considering that the constraints on distinct decays generally vary, new bounds can be established due to this mixing.

Additionally, in the $M_1^{-} \to M_2^{+} \ell_1^- \ell_2^-$ and $\tau^-\to \ell^+ M_1^-M_2^{-}$ decays, the matrix elements of operators that contain a tensor quark current (i.e., $\mathcal{O}_{2,4,7,8}$) vanish at first order after hadronization \cite{Quintero:2016iwi,deGouvea:2007qla}, therefore these operators are not bounded at the tree-level.
However, given that QCD corrections can mix the tensor quark operators ($\mathcal{O}_{2,4,7,8}$) at the quark level, residual contributions from these WCs ($C_{2,4,7,8}$) appear in the QCD-corrected partial decay width formulas.
This means that not only improved limits on WCs are found by considering QCD corrections, but also new limits on tensor operators not constrained at the tree-level. 

\begin{figure}[t]
\centering
\includegraphics[width=0.35\linewidth]{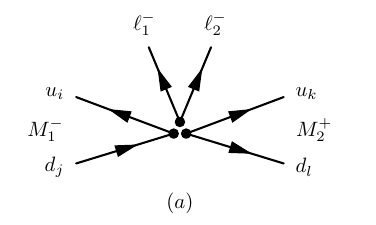}
\includegraphics[width=0.35\linewidth]{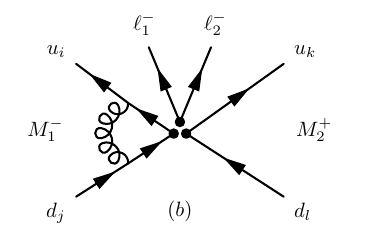}\\
\includegraphics[width=0.35\linewidth]{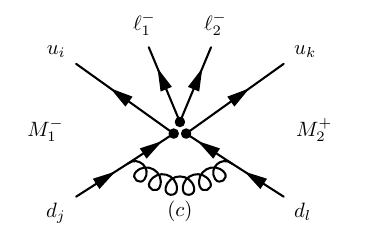}
\includegraphics[width=0.35\linewidth]{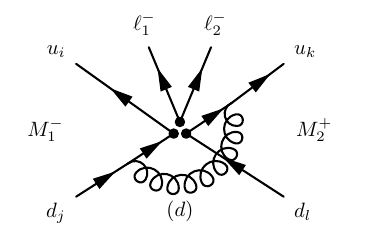}
\caption{Effective $d=9$ operator description of the short-range
  mechanisms (SRM) of the meson decay $M_1^{-} \to M_2^{+} \ell_1^- \ell_2^-$. Diagram (a) gives the tree-level
  description, while diagrams (b)-(d) are one-loop QCD corrections to the
amplitude.}\label{fig:shortM}
\end{figure}

\begin{figure}[tb]
\centering
\includegraphics[width=0.35\linewidth]{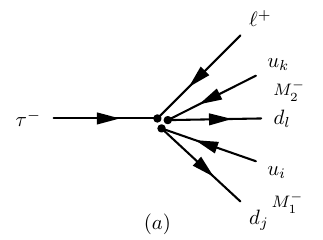}
\includegraphics[width=0.35\linewidth]{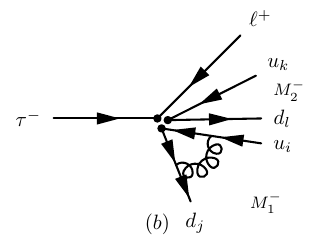}\\
\includegraphics[width=0.35\linewidth]{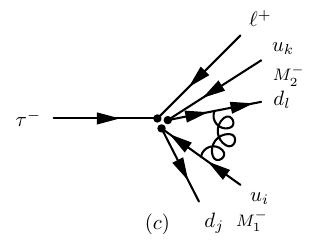}
\includegraphics[width=0.35\linewidth]{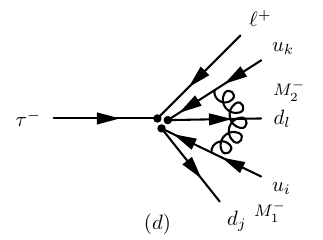}
\caption{Effective $d=9$ operator description of the short range
  mechanisms (SRM) of the decay $\tau^-\to \ell^+ M_1^-M_2^-$. Diagram (a) gives the tree-level
  description, while diagrams (b)-(d) are one-loop QCD corrections to the
amplitude.}\label{fig:shortTau}
\end{figure}

Figures (\ref{fig:shortM}) and (\ref{fig:shortTau}) show the one-loop QCD corrections applied to the meson and tau decays, respectively.\footnote{{In this work, we consider the RGE evolution driven by QCD corrections below the electroweak scale, specifically from $\Lambda = m_Z$ to $\mu = 1\,\text{GeV}$. Although electroweak corrections above the $m_Z$ scale, of order $\mathcal O(\alpha_2/4\pi)$, are beyond the scope of this study and are comparatively small relative to QCD corrections, they may still be of interest due to the additional mixing effects they can induce, particularly with operators relevant to $0\nu\beta\beta$ decay.}}
As it is well-known \cite{Gonzalez:2015ady}, operator mixing under renormalization appears given the existence of a non-diagonal renormalization matrix.
Equivalently, this mixing can be understood as a mixing between the WCs.

Using the renormalization group (RG) equation formalism outlined for $0\nu\beta\beta$-decay in Ref.~\cite{Gonzalez:2015ady}, the RG equation for the WCs is the following:
\begin{equation}\label{eq:RGE-1}
\frac{d \vec{C}(\mu)}{d \ln\mu} = \hat\gamma^{T}\vec{C}(\mu).
\end{equation}
Here, the vector $\vec{C} = (C_{1}, C_{2}, \ldots)$ contains the Wilson coefficients, and $\hat\gamma^{T}$ is the transposed anomalous dimension matrix.
The solution to Eq.~\eqref{eq:RGE-1} is given by:
\begin{equation}\label{eq:RGE-Sol}
\vec{C}(\mu) = \hat{U}(\mu, \Lambda) \cdot \vec{C}(\Lambda),
\end{equation}
where $\hat{U}(\mu, \Lambda)$ serves as an evolution matrix linking the high-energy scale $\Lambda$ and the low-energy scale $\mu$.

At leading order (LO), the evolution matrix is:
\begin{equation}\label{eq:EV}
\hat{U}(\mu, \Lambda) = \hat{V}\,\text{Diag} \left\{ \left( \frac{\alpha_s(\Lambda)}{\alpha_s(\mu)} \right)^{\hat \gamma / (2 \beta_0)} \right\} \hat{V}^{-1},
\end{equation}
where the LO QCD running coupling constant is:
\begin{equation}
\alpha_s(\mu) = \frac{\alpha_s(\Lambda)}{1 - \frac{\beta_0 \alpha_s(\Lambda)}{2 \pi} \log \left( \frac{\Lambda}{\mu} \right)}.
\end{equation}
Here, $\beta_0 = (33 - 2f)/3$, where $f$ denotes the number of quark flavors with masses $m_f < \mu$.
We normalize using the value of $\alpha_s$ at the scale of the $Z$-boson mass: $\alpha_s(\mu = m_z) = 0.118$~\cite{Workman:2022ynf}. The matrix $\hat{V}$ in Eq.~\eqref{eq:EV} is defined as:
\begin{equation}
\hat {\gamma}_D = \hat{V}^{-1} \hat{\gamma}^T \hat{V},
\end{equation}
where $\hat{\gamma}$ is the anomalous dimension matrix, whose matrix elements are:
\begin{equation}
(\hat \gamma)_{ij} = \frac{\alpha_s}{4\pi}\left[-2(\hat b_{ij} - 2C_F \delta_{ij})\right],
\end{equation}
with $C_F = \frac{N^2 - 1}{2N}$ the standard $SU(N)$ color factor and the  matrix $\hat b_{ij}$ is derived from the one-loop QCD corrections of the operator basis~[Eqs.\,\eqref{eq:OperBasis-1}-\eqref{eq:OperBasis-8}]. The general expression for the one-loop QCD-corrected operator matrix elements at LO is:
\begin{equation}
\langle \mathcal{O}_{i}\rangle^{(0)} = \left[\delta_{ij} + \frac{\alpha_s}{4\pi} b_{ij} \left( \frac{1}{\epsilon} + \ln \frac{\mu^2}{-p^2} \right) \right] \langle\mathcal{O}_{j}\rangle^{tree},
\end{equation}
where $\langle\mathcal{O}_{j}\rangle^{tree}$ are the tree-level operator matrix elements. For further details on the renormalization procedure for $d=9$ effective operators with two quark currents, the reader is referred to Ref. \cite{Gonzalez:2015ady}.

Since due to Fierz transformations RG evolution can mix operators with different quark flavor configurations, we introduce the following notation in order to specify the different mixings in the anomalous dimension matrices.
If $ijkl=1234$ are the quark flavor indices of the operator $\mathcal O_n$, then we define the following permutations of the quark flavor indices:
\begin{align}
    \mathcal O_{n(1234)}^{'XYZ} = \mathcal O_{n(1432)}^{XYZ},\ \ \ 
    \overline{\mathcal{O}}_{n(1234)}^{XYZ} = \mathcal O_{n(3412)}^{XYZ},\ \ \ 
    \overline{\mathcal{O'}}_{n(1234)}^{XYZ} = \mathcal O_{n(3214)}^{XYZ}.
    \label{eq:Oprimebar}
\end{align}
Therefore, for example, the anomalous dimension matrix $\hat \gamma_{13'}^{XYZ}(ijkl)$ mixes operators $\mathcal O_{1(ijkl)}^{XYZ}$ and $\mathcal{O}_{3(ijkl)}^{'XYZ}=\mathcal{O}_{3(ilkj)}^{XYZ}$.
Note that for operators with repeated $u$-type quarks ($i=k$), $\mathcal O = \overline{\mathcal{O}}'$ 
and $\mathcal{O}' = \overline{\mathcal{O}}$, while operators with repeated $d$-type quarks ($j=l$) satisfy  $\mathcal{O}=\mathcal{O}'$ and $\overline{\mathcal{O}}=\overline{\mathcal{O}}'$.
On the other hand, for operators involving four different quark flavors ($i\neq k$, $j\neq l$), the operators $\mathcal{O}$, $\mathcal{O}'$, $\overline{\mathcal{O}}$, and $\overline{\mathcal{O}}'$ are all different.

The anomalous dimension matrices are listed next. They are denoted by $\hat \gamma_{n_1n_2,\ldots}^{XYZ}(ijkl)$, where $n_1n_2,\ldots$ represent the operators $\mathcal O_{n_1},\mathcal O_{n_2},\ldots$ mixed by the anomalous dimension matrix, $XYZ$ the chiralities of the operators, and the indices $i,k=u,c$, and $j,l=d,s,b$ represent quark flavors.
We group the anomalous dimensions in four groups depending on the symmetry of their flavor indices ($ijkl$):  (i) Repeated up-flavor quarks, (ii) repeated down-flavor quarks, (iii) repeated up- and down-flavor quarks, (iv) all quark flavors different.
Unless explicitly indicated, in the matrices below the flavor indices ($ijkl$) are all different.
\paragraph{Repeated up-flavor quarks ($i=k$):}

\begin{align}
\hat{\gamma}^{XXZ}_{(12)}(ijil)
=-2\left(
\begin{array}{cc}
6 C_F-3& \frac{N-2}{4 N}\\
-\frac{12(2+N)}{N}&-3-2 C_F
\end{array}
\right),\ \ \
\hat{\gamma}_{(13)}^{(LR/RL)Z}(ijil)
=-2\left(
\begin{array}{cc}
6 C_F&0\\
-6& -\frac{3}{N} 
\end{array}
\right),\label{eq:gamma12XXidkd}
\end{align}
\begin{align}
\hat{\gamma}_{(13)}^{(LR/RL)Z}(ijil)
=-2\left(
\begin{array}{cc}
6 C_F&0\\
-6& -\frac{3}{N} 
\end{array}
\right),\ \ \ 
\hat{\gamma}_{(3)}^{XXZ}(ijil)=-2\left(-3+\frac{3}{N}\right)
\end{align}
\begin{align}
\hat{\gamma}_{(44^{\prime}55^{\prime})}^{XXZ}(ijil)
=-2\left(
\begin{array}{cccc}
-C_F & -\frac{3}{2} & - \frac{3i}{N} & - \frac{3i}{2}\\
-\frac{3}{2} & -C_F & - \frac{3 i}{2} & - \frac{3 i}{N}\\
 \frac{i}{N} & - \frac{i}{2} & 3 C_F & -\frac{3}{2}\\
- \frac{i}{2} &  \frac{i}{N} & -\frac{3}{2} & 3 C_F
\end{array}\right),
\end{align}
\begin{align}
\hat{\gamma}_{(45)}^{(LR/RL)Z}(ijil)=
-2\left(
\begin{array}{cc}
-\frac{3}{2}-C_F & \frac{3i(N+2)}{2N} \\
\frac{i(N-2)}{2N} & -\frac{3}{2}+3C_F 
\end{array}\right),
\end{align}
\begin{align}
\hat{\gamma}_{(6)}^{XXX}(ijil) =-2 \left(-1 + \frac{1}{N}\right),\ \ \ 
\hat{\gamma}_{(6)}^{LLR/RRL}(ijil) =-2 \left(1+\frac{1}{N}\right),
\end{align}
\begin{align}
\hat{\gamma}_{(67)}^{LRR/RLL}(ijil) =-2 \left(
\begin{array}{cc}
 -\frac{1}{N} & \frac{i}{2} \\
 0 & 2 C_F \\
\end{array}
\right),
\end{align}
\begin{align}
\hat{\gamma} _{(7\overline{7}8)}^{XXX}(ijil) =-2 \left(
\begin{array}{ccc}
 2 C_F-\frac{1}{N}-2 & -\frac{1}{N} & -\frac{i(8+N)}{8 N} \\
 -\frac{1}{N} & 2 C_F-\frac{1}{N}-2 &  \frac{i(8+N)}{8 N}\\
 0 & 0 & -2 C_F
\end{array}
\right).
\end{align}

\paragraph{Repeated down-flavor quarks ($j=l$):}

\begin{align}
\hat{\gamma}^{XXZ}_{(12)}(ijkj)
=-2\left(
\begin{array}{cc}
6 C_F-3& \frac{N-2}{4 N}\\
-\frac{12(2+N)}{N}&-3-2 C_F
\end{array}
\right),\label{eq:gamma12XXidkd2}
\end{align}
\begin{align}
\hat{\gamma}_{(13)}^{(LR/RL)Z}(ijkj)
=-2\left(
\begin{array}{cc}
6 C_F&0\\
-6& -\frac{3}{N} 
\end{array}
\right),\ \ \ 
\hat{\gamma}_{(3)}^{XXZ}(ijkj)
=-2\left(-3+\frac{3}{N}\right)
\end{align}
\begin{align}
\hat{\gamma}_{(45)}^{XXZ}(ijkj)
=-2\left(
\begin{array}{cc}
-\frac{3}{2}-C_F & -\frac{3i(N+2)}{2N} \\
-\frac{i(N-2)}{2N} &  -\frac{3}{2}+3C_F
\end{array}\right),
\end{align}
\begin{align}
\hat{\gamma} _{(4\overline{4}5\overline{5})}^{(LR/RL)Z}(ijkj)
=-2 \left(
\begin{array}{cccc}
 -C_F & -\frac{3}{2} & \frac{3 i}{N} & \frac{3 i}{2} \\
 -\frac{3}{2} & -C_F & \frac{3 i}{2} & \frac{3 i}{N} \\
 -\frac{i}{N} & \frac{i}{2} & 3 C_F & -\frac{3}{2} \\
 \frac{i}{2} & -\frac{i}{N} & -\frac{3}{2} & 3 C_F
\end{array}
\right),
\end{align}
\begin{align}
\hat{\gamma}_{\left(6\right)}^{XXX}(ijkj) =-2\left( 1+\frac{1}{N}\right),\ \ \ 
\hat{\gamma}_{\left(6\right)}^{LLR/RRL}(ijkj) =-2 \left(
-1+\frac{1}{N}
\right),
\end{align}
\begin{align}
\hat{\gamma}_{\left(6\overline{7}\right)}^{LRR/RLL}(ijkj) =-2 \left(
\begin{array}{cc}
 -\frac{1}{N} & \frac{i}{2} \\
   0 & 2 C_F \\
\end{array}
\right),
\end{align}
\begin{align}
\hat{\gamma}_{\left(7\overline{7}8\right)}^{XXX}(ijkj) =-2 \left(
\begin{array}{ccc}
 2 C_F-\frac{1}{N} & -2-\frac{1}{N} & \frac{i(N-8)}{8N} \\
 -2-\frac{1}{N} & 2 C_F-\frac{1}{N} & -\frac{i(N-8)}{8N}  \\
 0 & 0 & -2 C_F 
\end{array}
\right).
\end{align}

\noindent
\textbf{Repeated up- and down-flavor quarks ($i=k$, $j=l$):}\footnote{This case, excepting for the possibility of different lepton flavors, is similar to the RG evolution of the operators involved in the short-range mechanism of $0\nu\beta\beta$-decay. This is relevant for the decays $\tau^-\to \ell^+ M_1^- M_2^-$, with $M_1^-=M_2^-$.}

\begin{align}
\hat{\gamma}^{XXZ}_{(12)}(ijij)=-2\left(
\begin{array}{cc}
6 C_F-3& \frac{N-2}{4 N}\\
-\frac{12(2+N)}{N}&-3-2 C_F
\end{array}
\right),\ \ \ 
\hat{\gamma}_{(13)}^{(LR/RL)Z}(ijij)=-2\left(
\begin{array}{cc}
6 C_F&0\\
-6& -\frac{3}{N} 
\end{array}
\right),
\end{align}
\begin{align}
\small
\hat{\gamma}_{(3)}^{XXZ}(ijij)=-2\left(-3+\frac{3}{N}\right),\ \ \ 
\hat{\gamma}_{(45)}^{XXZ}(ijij)=-2\left(
\begin{array}{cc}
-\frac{3}{2}-C_F &  -\frac{3 i(N+2)}{2N} \\
  -\frac{i(N-2)}{2 N} &  -\frac{3}{2}+3 C_F \\
\end{array}
\right),
\normalsize
\end{align}
\begin{align}
\hat{\gamma}_{(45)}^{(LR/RL)Z}(ijij)=-2 \left(
\begin{array}{cc}
-\frac{3}{2}-C_F &  \frac{3 i(N+2)}{2N} \\
  \frac{i(N-2)}{2 N} &  -\frac{3}{2}+3 C_F \\
\end{array}
\right),
\end{align}
\begin{align}
\hat{\gamma}_{(67)}^{LRR/RLL}(ijij)=-2 \left(
\begin{array}{cc}
 -\frac{1}{N} & \frac{i}{2} \\
 0 & 2 C_F \\
\end{array}
\right),
\end{align}
\begin{align}
    \hat{\gamma}_{(7)}^{XXX}(ijij)=-2\left(-2+2 C_F-\frac{2}{N}\right)= 0,\mbox{   (for $N=3$).}\label{eq:gamma7XXXudud}
\end{align}

\paragraph{All quark flavors different:}

\begin{align}
\hat{\gamma}^{XXZ}_{(11^{\prime}22^{\prime})}(ijkl) 
=-2\left(
\begin{array}{cccc}
6 C_F & -3 & -\frac{1}{2N} & \frac{1}{4} \\
-3 & 6 C_F &  \frac{1}{4} & -\frac{1}{2N}\\
-\frac{24}{N} & -12 & -2 C_F &-3\\
-12 & -\frac{24}{N} & -3 & -2 C_F
\end{array}\right),
\end{align}
\begin{align}
\hat{\gamma}_{(13^{\prime})}^{(LR/RL)Z} (ijkl)
=-2\left(
\begin{array}{cc}
6 C_F&0\\
-6& -\frac{3}{N} 
\end{array}
\right),\ \ \ 
\hat{\gamma}_{(33^{\prime})}^{XXZ}(ijkl)=-2\left(
\begin{array}{cc}
 \frac{3}{N}& -3  \\
 -3 & \frac{3}{N}
\end{array}\right),
\end{align}
\begin{align}
\hat{\gamma}_{(44^{\prime}55^{\prime})}^{XXZ}(ijkl)
=-2\left(
\begin{array}{cccc}
-C_F & -\frac{3}{2} & - \frac{3i}{N} & - \frac{3 i}{2}\\
-\frac{3}{2} & -C_F & - \frac{3 i}{2} & - \frac{3 i}{N}\\
 \frac{i}{N} & - \frac{i}{2} & 3 C_F & -\frac{3}{2}\\
- \frac{i}{2} &  \frac{i}{N} & -\frac{3}{2} & 3 C_F
\end{array}\right),
\end{align}
\begin{align}
\hat{\gamma} _{(4\overline{4}'5\overline{5}')}^{(LR/RL)Z}(ijkl)
=-2 \left(
\begin{array}{cccc}
 -C_F & -\frac{3}{2} & \frac{3 i}{N} & \frac{3 i}{2} \\
 -\frac{3}{2} & -C_F & \frac{3 i}{2} & \frac{3 i}{N} \\
 -\frac{i}{N} & \frac{i}{2} & 3 C_F & -\frac{3}{2} \\
 \frac{i}{2} & -\frac{i}{N} & -\frac{3}{2} & 3 C_F
\end{array}
\right),
\end{align}
\begin{align}
\hat{\gamma}_{\left(66'\right)}^{XXX}(ijkl) =-2\left( 
\begin{array}{cc}
 \frac{1}{N} & 1 \\
 1 & \frac{1}{N} \\
\end{array}
\right),\ \ \ 
\hat{\gamma} _{\left(66'\right)}^{LLR/RRL}(ijkl) =-2\left( 
\begin{array}{cc}
 \frac{1}{N} & -1 \\
 -1 & \frac{1}{N} \\
\end{array}
\right),    
\end{align}
\begin{align}
\hat{\gamma}_{\left(6\overline{7'}\right)}^{LRR/RLL}(ijkl) =-2 \left(
\begin{array}{cc}
 -\frac{1}{N} & \frac{i}{2} \\
   0 & 2 C_F \\
\end{array}
\right),
\end{align}
\begin{align}
\hat{\gamma}_{\left(7 7' \overline{7}\overline{7}'88'\right)}^{XXX}(ijkl) =-2 \left(
\begin{array}{cccccc}
 2 C_F-\frac{1}{N} & 0 & -\frac{1}{N} & -2 & -\frac{i}{N} & \frac{i}{8} \\
 0 & 2 C_F-\frac{1}{N} & -2 & -\frac{1}{N} & \frac{i}{8} & -\frac{i}{N} \\
 -\frac{1}{N} & -2 & 2 C_F-\frac{1}{N} & 0 & \frac{i}{N} & -\frac{i}{8} \\
 -2 & -\frac{1}{N} & 0 & 2 C_F - \frac{1}{N} & -\frac{i}{8} & \frac{i}{N}  \\
 0 & 0 & 0 & 0 & -2 C_F & 0\\
 0 & 0 & 0 & 0 & 0 & -2 C_F
\end{array}\label{eq:gamma78xxxijkl}
\right).
\end{align}

From the previous anomalous dimension matrices, Eqs.\,(\ref{eq:gamma12XXidkd})-(\ref{eq:gamma78xxxijkl}) and the definition of $\hat U(\mu,\Lambda)$  in Eq.\,(\ref{eq:EV}), we obtain the $\mu$-evolution matrices for the RG evolution from $\Lambda=m_Z$ to $\mu=1\,\mbox{GeV}$.
It is important to account for the quark thresholds when considering the evolution of the Wilson coefficients. 
To address this, we consider consecutive $\mu$-evolution matrices, each corresponding to a different number ($f$) of quark flavors:

\begin{equation}
\hat{U}(\mu, \Lambda = m_Z) = \hat{U}^{(f=3)}(\mu, \mu_c) \hat{U}^{(f=4)}(\mu_c, \mu_b) \hat{U}^{(f=5)}(\mu_b, m_Z),\label{eq:UmuLambda}
\end{equation}
where the intermediate scales are the corresponding quark masses ($\mu_c=m_c$, $\mu_b=m_b$).
This evolution matrices will be used in the next section and are listed in Appendix (\ref{app:umatrices}).

\section{RGE-Induced Bounds on Wilson Coefficients}\label{sec:RGEinducedbounds}

In this section, we derive RGE-improved bounds on the Wilson coefficients ($C_{n(ijkl)}^{XYZ}$), which are associated with the dimension-9 operator basis listed in Eqs.\,(\ref{eq:OperBasis-1})-(\ref{eq:OperBasis-8}).
These operators contribute to the LNV meson decays $M_1^-\to M_2^+ \ell_1^- \ell_2^-$ and to the LNV tau decays $\tau^-\to \ell^+ M_1^- M_2^-$.
Many of these processes have been searched experimentally, and their non-observation puts upper bounds on their branching fractions.
The existing experimental limits on the branching fractions for different LNV meson and tau decay are shown in Tables (\ref{tab:expbounds_mesondecays}) and (\ref{tab:expbounds_taudecays}). 

Using the expressions for the partial widths given in Eqs. (\ref{eq:mesondecaywidth}) and (\ref{eq:taudecaywidth}) for meson and tau decays in terms of the Wilson coefficients, we can translate the experimental bounds listed in Tables (\ref{tab:expbounds_mesondecays}) and (\ref{tab:expbounds_taudecays}) into constraints on the respective Wilson coefficients.
The resultant bounds are presented in Tables (\ref{tab:limits_mesons_O12_ee_noCKMbasis})-(\ref{tab:limits_taus_7_noCKMbasis}) under the ``Tree Level'' column, as they do not take into account RG evolution effects.
{We can see that the most stringent tree-level bounds on the Wilson coefficients stem from the LNV kaon and $B$-meson decays $K^-\to\pi^+\ell_1^-\ell_2^-$ and $B^-\to \pi^+\mu^-\mu^-$ [of order $\mathcal O(1)$ and $\mathcal O(10^2)$ respectively] since these processes have the strongest experimental bounds (see Table \ref{tab:expbounds_mesondecays}). The remaining processes translate into tree-level bounds of the Wilson coefficients up to three orders of magnitude weaker.\footnote{{This range of bounds on the Wilson coefficients ($C_n$) translated into the new physics scale $\Lambda_{\text{NP}}$ (through $1/\Lambda_{\text{NP}}^5 = 
 C_n G_F^2/2m$) corresponds to $\Lambda_{\text{NP}}\gtrsim (8-94)\,\mbox{GeV}$.}}
}

The RG evolution matrices $U(\mu,\Lambda)$, calculated in Sec.\,(\ref{sec:QCDcorrections}) and listed in Appendix (\ref{app:umatrices}), allow us to derive new limits on the Wilson coefficients.
If at the tree level the Wilson coefficient $C_n$ appearing in the expressions for the partial decay widths [Eqs.\,(\ref{eq:mesondecaywidth}) and (\ref{eq:taudecaywidth})] has a tree-level bound $C_n \lesssim C_n^{exp}$, then
due to radiative corrections, the tree-level Wilson coefficient $C_n$ should be replaced by $C_n(\mu)$ according to Eq.\,(\ref{eq:RGE-Sol}), i.e,
\begin{align}
C_n \to C_n(\mu) = \sum_{m}\hat U_{nm}(\mu,\Lambda) C_m(\Lambda) \lesssim C_n^{exp},\label{eq:newbounds}
\end{align}
where the indices $n,m$ run over the operators that are mixed by the evolution matrix $\hat U(\mu,\Lambda)$.
In order to extract the RGE improved bounds from Eq.\,(\ref{eq:newbounds}), we apply the conventional assumption about the presence of only one operator at a time in the sum in Eq.\,(\ref{eq:newbounds}).
These simplified ``\textit{on-axis}'' limits on $C_n(\Lambda)$ are shown in Tables (\ref{tab:limits_mesons_O12_ee_noCKMbasis})-(\ref{tab:limits_taus_7_noCKMbasis}) under the columns labeled ``With QCD Corrections''.
The numerical values of the constants used in the calculations are listed in Appendix (\ref{app:constants}).

The most notable results are the new induced bounds on operators that are not constrained at the tree level.
This is the case for operators $\mathcal O_2$, $\mathcal O_4$, and $\mathcal O_7$, whose respective RGE-induced bounds are shown in gray-color background in Tables (\ref{tab:limits_mesons_O12_ee_noCKMbasis}), (\ref{tab:limits_mesons_O45_ee_noCKMbasis}), and  (\ref{tab:limits_mesons_7_emu_noCKMbasis}) for meson decays, and in Tables (\ref{tab:limits_taus_O12}), (\ref{tab:limits_taus_O45}), and (\ref{tab:limits_taus_7_noCKMbasis}) for tau decays.
It can be seen from the evolution matrices in Appendix (\ref{app:umatrices}) that these new bounds on 
$C_2$, $C_4$, and $C_7$, 
are inherited due to mixing from the tree-level bounds on 
$C_1$, $C_5$, and $C_6$, 
respectively. 

In Sec.\,(\ref{sec:QCDcorrections}), we found that the RG evolution for operators $\mathcal O_{n(ijkl)}^{XYZ}$ with different flavor quarks ($i\neq j\neq k\neq l$) also induces mixing with operators with different quarks flavor configurations 
(i.e., the $\mathcal O'$, $\mathcal{\overline{O}}$, and $\mathcal{\overline{O'}}$ operators defined in Eq.\,(\ref{eq:Oprimebar})).
These mixings also induce bounds on operators not constrained at the tree level, as is the case of the bounds coming from the processes $B^-\to D^+ \ell_1^-\ell_2^-$ and $B^-\to D_s^+ \mu^-\mu^-$. At the tree level, these processes constrain the operators $\mathcal O_{n(ubcd)}^{XYZ(\ell_1\ell_2)}$ and $\mathcal O_{n(ubcs)}^{XYZ(\mu\mu)}$, respectively, while due to RGE effects they also induce bounds on $\mathcal O_{n(udcb)}^{XYZ(\ell_1\ell_2)}$ and $\mathcal O_{n(uscb)}^{XYZ(\mu\mu)}$, as shown in Tables (\ref{tab:limits_mesons_O12_ee_noCKMbasis})-(\ref{tab:limits_mesons_6_emu_noCKMbasis}).

Note that the operators bounded by the processes $D^-\to K^+ \ell_1^-\ell_2^-$ and $D_s^-\to \pi^+ \ell_1^-\ell_2^-$ mix under RG evolution, so these processes put bounds on the same set of operators. For this reason,  when we inform the bounds on the Wilson coefficients associated to these processes, for comparison purposes we use the average mass of the $D^-$ and $D_s^-$ mesons for the mass scale $m$ appearing in the definition of the Wilson coefficient [see Eq.\,(\ref{eq:effL})].

Additionally, we find that the bounds on the Wilson coefficients $C_1^{XXZ}$ and $C_1^{(LR,RL)Z}$ are improved by a factor of $\sim 2$ and $\sim 3$, respectively, as can be seen in Tables (\ref{tab:limits_mesons_O12_ee_noCKMbasis}) and (\ref{tab:limits_taus_O12}).

\section{Summary and Conclusions}\label{sec:conclusions}

In this work, we calculate RGE-improved bounds on the Wilson coefficients of the LNV ($\Delta L = 2$) low energy effective Lagrangian 
presented in Sec.\,(\ref{sec:LNVLag}).
A subset of these operators, particularly $\mathcal O_{1,2,3,4,5}$ with first-generation quarks and leptons, contributes to the short-range mechanism of $0\nu\beta\beta$-decay, and therefore they are strongly constrained.
Operators with other quark or lepton flavor content (which are not constrained by $0\nu\beta\beta$-decay) are bounded by the non-observation of different LNV meson and tau decays.

We calculate QCD corrections to this operator basis and find the RG evolution matrices, which show how the Wilson coefficients evolve from one energy scale to another.
Unlike the $0\nu\beta\beta$ case studied in the literature, here the general flavor structure results in the mixing of not only different Lorentz structures but also distinct quark flavor configurations.
This implies that after hadronization there can be mixing between operators responsible for the decay of distinct mesons, leading to new bounds on certain operators (e.g., the operators contributing to $D^-\to K^+ \ell^-\ell^-$ and $D_s^- \to \pi^+ \ell^-\ell^-$).

We use the existing experimental limits on 3-body LNV meson and tau decays [shown in Tables (\ref{tab:expbounds_mesondecays}) and (\ref{tab:expbounds_taudecays})] and the expressions for their partial decay widths in terms of the corresponding Wilson coefficients, to convert these experimental bounds into constraints on the Wilson coefficients. Then we use the RG evolution matrices calculated in section (\ref{sec:QCDcorrections}) to derive QCD-improved bounds on the Wilson coefficients. Tables (\ref{tab:limits_mesons_O12_ee_noCKMbasis})-(\ref{tab:limits_taus_7_noCKMbasis}) show our results under the column ``With QCD Corrections''. For comparison, we include the tree-level results under the column ``Tree Level''.

One of the significant findings are the new constraints on certain operators that were not bounded at the tree level. This was the case for operators $\mathcal O_2$, $\mathcal O_4$, and $\mathcal O_7$.
Furthermore, the RG evolution led to the mixing of certain operators with operators of a different quark-flavor configuration. As a result, new constraints appear on operators that were not initially constrained, as is the case for operators contributing to $B^-\to D^+ \ell_1^-\ell_2^-$ and $B^-\to D_s^+ \mu^-\mu^-$.
Additionally, we found that RG evolution improves the bounds on the Wilson coefficients $C_1^{XXZ}$ and $C_1^{(LR/RL)Z}$ by factors of 2 and 3, respectively,
excepting for the operators contributing $D^-\to K^+ \ell_1^-\ell_2^-$ and $D_s^- \to \pi^+ \ell_1^-\ell_2^-$, where the improvement can be up to a factor of $20$.

\begin{table}[b!]\footnotesize
\centering
\begin{NiceTabular}{c|c|cc|ccc}[colortbl-like]
\multicolumn{2}{c}{ }  & \multicolumn{2}{c}{Tree Level} & \multicolumn{3}{c}{With QCD Corrections}\\
\hline
Process &$(ij)(kl)(\ell_1\ell_2)$ & 
$\left|C_{1(ijkl)}^{XYZ}\right|$ & $\left|C_{2(ijkl)}^{XXX}\right|$ &  $\left|C_{1(ijkl)}^{XXZ}\right|$ & $\left|C_{1(ijkl)}^{(LR,RL)Z}\right|$ & $\left|C_{2(ijkl)}^{XXX}\right|$ \\
\hline
\hline
$K^-\to \pi^+ e^-e^-$ & $(us)(ud)(ee)$ &
$2.0$ & $-$ & 
\impb{$9.9\times 10^{-1}$} & 
\impb{$6.5\times 10^{-1}$} & 
\newb{$2.0\times 10^{2}$} \\ 
\hline
$D^-\to \pi^+ e^-e^-$ & $(cd)(ud)(ee)$ &
$4.8\times 10^3$ & $-$ & 
\impb{$2.4\times 10^{3}$} & 
\impb{$1.6\times 10^{3}$} & 
\newb{$4.8\times 10^{5}$} \\ 
\hline
%
$D^-\to K^+ e^-e^-$ & $(cd)(us)(ee)$ & 
$9.3\times 10^{3}$ & $-$ &
\impb{$2.8\times 10^{3}$} & 
\impb{$3.1\times 10^{3}$} & 
\newb{$1.5\times 10^{5}$} \\ 
%
$D_s^-\to \pi^+ e^-e^-$  & $(cs)(ud)(ee)$ & 
$8.3\times 10^3$ & $-$ &
\impb{$2.5\times 10^{3}$} & 
\impb{$2.7\times 10^{3}$} & 
\newb{$1.3\times 10^{5}$} \\ 
\hline

{$D_s^-\to K^+ e^-e^-$} & $(cs)(us)(ee)$ & 
$8.9\times 10^{3}$ & $-$ & 
\impb{$4.5\times 10^{3}$} & 
\impb{$2.9\times 10^{3}$} & 
\newb{$1.2\times 10^{5}$} \\ 
\hline

{$B^-\to \pi^+ e^-e^-$} & $(ub)(ud)(ee)$ & 
$2.0\times 10^{2}$ & $-$ &
\impb{$1.0\times 10^{2}$} & 
\impb{$6.7\times 10^{1}$} & 
\newb{$2.0\times 10^{4}$} \\ 
\hline

$B^-\to K^+ e^-e^-$ & $(ub)(us)(ee)$ & 
$2.4\times 10^{2}$ & $-$ &
\impb{$1.2\times 10^{2}$} & 
\impb{$7.8\times 10^{1}$} & 
\newb{$2.4\times 10^{4}$} \\ 
\hline

\multirow{2}*{$B^-\to D^+ e^-e^-$} & $(ub)(cd)(ee)$ & 
$2.9\times 10^{3}$ & $-$ & 
\impb{$8.8\times 10^{2}$} & 
\impb{$9.5\times 10^{2}$} & 
\newb{$4.8\times 10^{4}$} \\ 
 & $(ud)(cb)(ee)$ & 
$-$ & $-$ &
\newb{$2.2\times 10^{3}$} & 
\newb{$2.9\times 10^{3}$} & 
\newb{$4.1\times 10^{4}$} \\ 
\hline
\hline

{$K^-\to \pi^+ \mu^-\mu^-$} & $(us)(ud)(\mu\mu)$ & 
$2.9$ & $-$ &
\impb{$1.5$} & 
\impb{$9.7\times 10^{-1}$} & 
\newb{$2.9\times 10^{2}$} \\ 
\hline
%
%
{$D^-\to \pi^+ \mu^-\mu^-$} & $(cd)(ud)(\mu\mu)$ & 
$7.9\times 10^{2}$ & $-$ & 
\impb{$4.0\times 10^{2}$} & 
\impb{$2.6\times 10^{2}$} & 
\newb{$7.9\times 10^{4}$} \\ 
\hline
%
%
{$D^-\to K^+ \mu^-\mu^-$} & $(cd)(us)(\mu\mu)$ & 
$3.2\times 10^{4}$ & $-$ &
\impb{$1.6\times 10^{3}$} & 
\impb{$1.1\times 10^{4}$} & 
\newb{$3.0\times 10^{4}$} \\ 
$D_s^-\to \pi^+ \mu^-\mu^-$ & $(cs)(ud)(\mu\mu)$ & 
$2.1\times 10^3$ & $-$ &
\impb{$6.4\times 10^{2}$} & 
\impb{$6.9\times 10^{2}$} & 
\newb{$3.5\times 10^{4}$} \\ 
\hline
%
%
%
{$D_s^-\to K^+ \mu^-\mu^-$} & $(cs)(us)(\mu\mu)$ & 
$1.7\times 10^{3}$ & $-$ &
\impb{$8.4\times 10^{2}$} & 
\impb{$5.5\times 10^{2}$} & 
\newb{$1.7\times 10^{5}$} \\ 
\hline
%
%
{$B^-\to \pi^+ \mu^-\mu^-$} & $(ub)(ud)(\mu\mu)$ & 
$8.5\times 10^{1}$ & $-$ & 
\impb{$4.3\times 10^{1}$} & 
\impb{$2.8\times 10^{1}$} & 
\newb{$8.5\times 10^{3}$} \\ 
\hline
%
%
{$B^-\to K^+ \mu^-\mu^-$} & $(ub)(us)(\mu\mu)$ & 
$2.8\times 10^{2}$ & $-$ & 
\impb{$1.4\times 10^{2}$} & 
\impb{$9.2\times 10^{1}$} & 
\newb{$2.8\times 10^{4}$} \\ 
\hline
%
%
\multirow{2}*{$B^-\to D^+ \mu^-\mu^-$} & $(ub)(cd)(\mu\mu)$ & 
$1.5\times 10^{3}$ & $-$ &
\impb{$4.6\times 10^{2}$} & 
\impb{$4.9\times 10^{2}$} & 
\newb{$2.5\times 10^{4}$} \\ 
 & $(ud)(cb)(\mu\mu)$ & 
$-$ & $-$ & 
\newb{$1.2\times 10^{3}$} & 
\newb{$1.5\times 10^{3}$} & 
\newb{$2.1\times 10^{4}$} \\ 
\hline
%
%
\multirow{2}*{$B^-\to D_s^+ \mu^-\mu^-$} & $(ub)(cs)(\mu\mu)$ & 
$1.2\times 10^{3}$ & $-$ &
\impb{$3.7\times 10^{2}$} & 
\impb{$3.9\times 10^{2}$} & 
\newb{$2.0\times 10^{4}$} \\ 
 & $(us)(cb)(\mu\mu)$ & 
$-$ & $-$ &
\newb{$9.3\times 10^{2}$} & 
\newb{$1.2\times 10^{3}$} & 
\newb{$1.7\times 10^{4}$} \\ 
\hline
\hline
{$K^-\to \pi^+ e^-\mu^-$} & $(us)(ud)(e\mu)$ & 
$1.5$ & $-$ &
$7.7\times 10^{-1}$ & 
\impb{$5.1\times 10^{-1}$} & 
\newb{$1.5\times 10^{2}$} \\ 
\hline
%
%
{$D^-\to \pi^+ e^-\mu^-$} & $(cd)(ud)(e\mu)$ & 
$1.7\times 10^{3}$ & $-$ &
\impb{$8.5\times 10^{2}$} & 
\impb{$5.6\times 10^{2}$} & 
\newb{$1.7\times 10^{5}$} \\ 
\hline
%
{$D^-\to K^+ e^-\mu^-$} & $(cd)(us)(e\mu)$ & 
$9.7\times 10^{3}$ & $-$ &
\impb{$3.0\times 10^{3}$} & 
\impb{$3.2\times 10^{3}$} & 
\newb{$5.7\times 10^{4}$} \\ 
$D_s^-\to \pi^+ e^-\mu^-$ & $(cs)(ud)(e\mu)$ & 
$4.0\times 10^3$ & $-$ &
\impb{$1.2\times 10^{3}$} & 
\impb{$1.3\times 10^{3}$} & 
\newb{$6.6\times 10^{4}$} \\ 
\hline
%
%
%
{$D_s^-\to K^+ e^-\mu^-$} & $(cs)(us)(e\mu)$ & 
$3.7\times 10^{3}$ & $-$ &
\impb{$1.9\times 10^{3}$} & 
\impb{$1.2\times 10^{3}$} & 
\newb{$3.7\times 10^{5}$} \\ 
\hline
%
%
{$B^-\to \pi^+ e^-\mu^-$} & $(ub)(ud)(e\mu)$ & 
$3.7\times 10^{2}$ & $-$ &
\impb{$1.9\times 10^{2}$} & 
\impb{$1.2\times 10^{2}$} & 
\newb{$3.7\times 10^{4}$} \\ 
\hline
%
%
{$B^-\to K^+ e^-\mu^-$} & $(ub)(us)(e\mu)$ & 
$3.9\times 10^{2}$ & $-$ &
\impb{$2.0\times 10^{2}$} & 
\impb{$1.3\times 10^{2}$} & 
\newb{$3.9\times 10^{4}$} \\ 
\hline
%
%
\multirow{2}*{$B^-\to D^+ e^-\mu^-$} & $(ub)(cd)(e\mu)$ & 
$1.7\times 10^{3}$ & $-$ & 
\impb{$5.2\times 10^{2}$} & 
\impb{$5.6\times 10^{2}$} & 
\newb{$2.8\times 10^{4}$} \\ 
 & $(ud)(cb)(e\mu)$ & 
$-$ & $-$ & 
\newb{$1.3\times 10^{3}$} & 
\newb{$1.7\times 10^{3}$} & 
\newb{$2.4\times 10^{4}$} \\ 
\hline
\end{NiceTabular}
\caption{Upper bounds on the Wilson coefficients $C_{1(ijkl)}^{XYZ(\ell_1\ell_2)}$ and  $C_{2(ijkl)}^{XXX(\ell_1\ell_2)}$ extracted from the experimental limits on the LNV meson decays $M_1\to M_2\, \ell_1^- \ell_2^-$ shown in Table (\ref{tab:expbounds_mesondecays}), with and without QCD corrections.
The bounds in gray background are new results present only after including the effects of QCD corrections.
The bounds which improve by at least a factor of 2 with respect to the corresponding tree-level bound are shown in light-gray.
Note that the operators bounded by the processes $D^-\to K^+ \ell_1^-\ell_2^-$ and $D_s^-\to \pi^+ \ell_1^-\ell_2^-$ mix under RG evolution, so these processes put bounds on the same sets of operators.
}
\label{tab:limits_mesons_O12_ee_noCKMbasis}
\end{table}
\begin{table}[h!]\footnotesize
\centering
\begin{NiceTabular}{c|c|c|cc}[colortbl-like]
\multicolumn{2}{c}{ }  & \multicolumn{1}{c}{Tree Level} & \multicolumn{2}{c}{With QCD Corrections}\\
\hline
Process &$(ij)(kl)(\ell_1\ell_2)$ & 
 $\left|C_{3(ijkl)}^{XYZ}\right|$ &  $\left|C_{3(ijkl)}^{XXZ}\right|$ & $\left|C_{3(ijkl)}^{(LR,RL)Z}\right|$
\\
\hline
\hline
$K^-\to \pi^+ e^-e^-$ & $(us)(ud)(ee)$ &
$1.5\times 10^{2}$ &
$1.9\times 10^{2}$ & 
$1.7\times 10^{2}$\\ 
\hline
$D^-\to \pi^+ e^-e^-$ & $(cd)(ud)(ee)$ &
$3.7\times 10^4$ &
$4.8\times 10^{4}$ & 
$4.2\times 10^{4}$\\ 
\hline
%
$D^-\to K^+ e^-e^-$ & $(cd)(us)(ee)$ & 
$4.8\times 10^{4}$ &
$3.8\times 10^{4}$ & 
$5.5\times 10^{4}$\\ 
%
$D_s^-\to \pi^+ e^-e^-$  & $(cs)(ud)(ee)$ & 
$6.0\times 10^4$ &
$4.8\times 10^{4}$ & 
$6.9\times 10^4$\\ 
\hline

{$D_s^-\to K^+ e^-e^-$} & $(cs)(us)(ee)$ & 
$4.4\times 10^{4}$ &
$5.7\times 10^{4}$ & 
$5.0\times 10^{4}$\\ 
\hline

{$B^-\to \pi^+ e^-e^-$} & $(ub)(ud)(ee)$ & 
$5.0\times 10^{2}$ &
$6.6\times 10^{2}$ & 
$5.8\times 10^{2}$\\ 
\hline

$B^-\to K^+ e^-e^-$ & $(ub)(us)(ee)$ & 
$4.8\times 10^{2}$ &
$6.4\times 10^{2}$ & 
$5.6\times 10^{2}$\\ 
\hline

\multirow{2}*{$B^-\to D^+ e^-e^-$} & $(ub)(cd)(ee)$ & 
$4.2\times 10^{3}$ &
$3.4\times 10^{3}$ & 
$4.8\times 10^{3}$\\ 
 & $(ud)(cb)(ee)$ & 
$-$ &
\newb{$8.6\times 10^{3}$} & 
$-$\\ 
\hline
\hline

{$K^-\to \pi^+ \mu^-\mu^-$} & $(us)(ud)(\mu\mu)$ & 
$2.4\times 10^{2}$ &
$3.2\times 10^{2}$ & 
$2.8\times 10^{2}$\\ 
\hline
%
%
{$D^-\to \pi^+ \mu^-\mu^-$} & $(cd)(ud)(\mu\mu)$ & 
$6.2\times 10^{3}$ &
$8.1\times 10^{3}$ & 
$7.1\times 10^{3}$\\ 
\hline
%
%
{$D^-\to K^+ \mu^-\mu^-$} & $(cd)(us)(\mu\mu)$ & 
$1.7\times 10^{5}$ &
\impb{$3.3\times 10^{4}$} & 
$1.9\times 10^{5}$\\ 
$D_s^-\to \pi^+ \mu^-\mu^-$ & $(cs)(ud)(\mu\mu)$ & 
$1.5\times 10^4$ &
$1.2\times 10^{4}$ & 
$1.8\times 10^4$\\ 
\hline
%
%
%
{$D_s^-\to K^+ \mu^-\mu^-$} & $(cs)(us)(\mu\mu)$ & 
$8.3\times 10^{3}$ &
$1.1\times 10^{4}$ & 
$9.5\times 10^{3}$\\ 
\hline
%
%
{$B^-\to \pi^+ \mu^-\mu^-$} & $(ub)(ud)(\mu\mu)$ & 
$2.1\times 10^{2}$ &
$2.8\times 10^{2}$ & 
$2.4\times 10^{2}$\\ 
\hline
%
%
{$B^-\to K^+ \mu^-\mu^-$} & $(ub)(us)(\mu\mu)$ & 
$5.7\times 10^{2}$ &
$7.5\times 10^{2}$ & 
$6.5\times 10^{2}$\\ 
\hline
%
%
\multirow{2}*{$B^-\to D^+ \mu^-\mu^-$} & $(ub)(cd)(\mu\mu)$ & 
$2.2\times 10^{3}$ &
$1.7\times 10^{3}$ & 
$2.5\times 10^{3}$\\ 
 & $(ud)(cb)(\mu\mu)$ & 
$-$ &
\newb{$4.5\times 10^{3}$} & 
$-$\\ 
\hline
%
%
\multirow{2}*{$B^-\to D_s^+ \mu^-\mu^-$} & $(ub)(cs)(\mu\mu)$ & 
$1.8\times 10^{3}$ &
$1.4\times 10^{3}$ & 
$2.0\times 10^{3}$\\ 
 & $(us)(cb)(\mu\mu)$ & 
$-$ &
\newb{$3.6\times 10^{3}$} & 
$-$\\ 
\hline
\hline
{$K^-\to \pi^+ e^-\mu^-$} & $(us)(ud)(e\mu)$ & 
$1.2\times 10^{2}$ &
$1.6\times 10^{2}$ & 
$1.4\times 10^{2}$\\ 
\hline
%
%
{$D^-\to \pi^+ e^-\mu^-$} & $(cd)(ud)(e\mu)$ & 
$1.3\times 10^{4}$ &
$1.7\times 10^{4}$ & 
$1.5\times 10^{4}$\\ 
\hline
%
{$D^-\to K^+ e^-\mu^-$} & $(cd)(us)(e\mu)$ & 
$5.0\times 10^{4}$ &
$4.0\times 10^{4}$ & 
$5.8\times 10^{4}$\\ 
$D_s^-\to \pi^+ e^-\mu^-$ & $(cs)(ud)(e\mu)$ & 
$2.9\times 10^4$ &
$2.3\times 10^{4}$ & 
$3.3\times 10^{4}$\\ 
\hline
%
%
%
{$D_s^-\to K^+ e^-\mu^-$} & $(cs)(us)(e\mu)$ & 
$1.8\times 10^{4}$ &
$2.4\times 10^{4}$ & 
$2.1\times 10^{4}$\\ 
\hline
%
%
{$B^-\to \pi^+ e^-\mu^-$} & $(ub)(ud)(e\mu)$ & 
$9.1\times 10^{2}$ &
$1.2\times 10^{3}$ & 
$1.0\times 10^{3}$\\ 
\hline
%
%
{$B^-\to K^+ e^-\mu^-$} & $(ub)(us)(e\mu)$ & 
$7.9\times 10^{2}$ &
$1.0\times 10^{3}$ & 
$9.1\times 10^{2}$\\ 
\hline
%
%
\multirow{2}*{$B^-\to D^+ e^-\mu^-$} & $(ub)(cd)(e\mu)$ & 
$2.5\times 10^{3}$ &
$2.0\times 10^{3}$ & 
$2.9\times 10^{3}$\\ 
 & $(ud)(cb)(e\mu)$ & 
$-$ &
\newb{$5.1\times 10^{3}$} & 
$-$\\ 
\hline
\end{NiceTabular}
\caption{Upper bounds on the Wilson coefficients $C_{3(ijkl)}^{XYZ(\ell_1\ell_2)}$ extracted from the experimental limits on the LNV meson decays $M_1\to M_2\, \ell_1^- \ell_2^-$ shown in Table (\ref{tab:expbounds_mesondecays}), with and without QCD corrections.
The bounds in gray background are new results present only after including the effects of QCD corrections.
For the operators shown in this table, existing tree-level bounds are not improved by QCD effects, excepting for $C_{3(cdus)}^{XXZ(\mu\mu)}$ (in light gray) which improves due to the mixing under RG evolution of the operators contributing to the processes $D^-\to K^+ \ell_1^-\ell_2^-$ and $D_s^-\to \pi^+ \ell_1^-\ell_2^-$.
}
\label{tab:limits_mesons_O123_ee_noCKMbasis}
\end{table}

\FloatBarrier
\begin{table}[h!]
\scriptsize
\centering
\begin{NiceTabular}{c|c|cc|cccc}[colortbl-like]
\multicolumn{2}{c}{ }  & \multicolumn{2}{c}{Tree Level} & \multicolumn{4}{c}{With QCD Corrections}\\
\hline
Process &$(ij)(kl)(\ell_1\ell_2)$ & $\left|C_{4(ijkl)}^{XYZ}\right|$ & $\left|C_{5(ijkl)}^{XYZ}\right|$ & $\left|C_{4(ijkl)}^{XXZ}\right|$ & $\left|C_{4(ijkl)}^{(LR,RL)Z}\right|$ & $\left|C_{5(ijkl)}^{XXZ}\right|$ & $\left|C_{5(ijkl)}^{(LR,RL)Z}\right|$
\\
\hline
\hline
\multirow{2}*{$K^-\to \pi^+ e^-e^-$} &
$(us)(ud)(ee)$ &
$-$ & 
$1.6\times 10^1$ & 
\newb{$1.8\times 10^2$} & 
\newb{$8.0\times 10^2$} & 
$9.1$ & 
$1.1\times 10^1$\\ 
 &
$(ud)(us)(ee)$ &
$-$ & 
$1.8\times 10^1$ & 
\newb{$1.6\times 10^2$} & 
\newb{$9.0\times 10^2$} & 
$1.0\times 10^1$ & 
$1.3\times 10^1$\\ 
\hline
%
%
\multirow{2}*{$D^-\to \pi^+ e^-e^-$} &
$(cd)(ud)(ee)$ &
$-$ & 
$1.3\times 10^{4}$ & 
\newb{$6.7\times 10^{5}$} & 
\newb{$1.4\times 10^{5}$} & 
$9.5\times 10^{3}$ & 
$7.6\times 10^{3} $\\ 
 &
$(ud)(cd)(ee)$ &
$-$ & 
$1.4\times 10^{4}$ & 
\newb{$7.0\times 10^{5}$} & 
\newb{$1.3\times 10^{5}$} & 
$9.9\times 10^{3}$ & 
$7.9\times 10^{3}$\\ 
\hline
%
%
%
%
%
 &
$(cs)(ud)(ee)$ &
$-$ & 
$2.3\times 10^{4}$ & 
\newb{$2.2\times 10^{5}$} & 
\newb{$2.0\times 10^{5}$} & 
$1.3\times 10^{4}$ & 
$1.3\times 10^{4} $\\ 
\
$D^-\to K^+ e^-e^-$ &
$(ud)(cs)(ee)$ &
$-$ & 
$2.3\times 10^{4}$ & 
\newb{$2.0\times 10^{5}$} & 
\newb{$2.2\times 10^{5}$} & 
$1.3\times 10^{4}$ & 
$1.3\times 10^{4} $\\ 
$D_s^-\to \pi^+ e^-e^-$ &
$(cd)(us)(ee)$ &
$-$ & 
$2.2\times 10^4$ & 
\newb{$2.3\times 10^{5}$} & 
\newb{$2.3\times 10^{5}$} & 
$1.2\times 10^{4}$ & 
$1.2\times 10^{4}$\\ 
 &
$(us)(cd)(ee)$ &
$-$ & 
$2.0\times 10^4$ & 
\newb{$2.3\times 10^{5}$} & 
\newb{$2.3\times 10^{5}$} & 
$1.2\times 10^{4}$ & 
$1.2\times 10^{4}$\\ 
\hline
%
%
%
%
%
\multirow{2}*{$D_s^-\to K^+ e^-e^-$} &
$(cs)(us)(ee)$ &
$-$ & 
$2.1\times 10^{4}$ & 
\newb{$1.0\times 10^{6}$} & 
\newb{$1.9\times 10^{5}$} & 
$1.5\times 10^{4}$ & 
$1.2\times 10^{4} $\\ 
 &
$(us)(cs)(ee)$ &
$-$ & 
$1.9\times 10^{4}$ & 
\newb{$9.4\times 10^{5}$} & 
\newb{$2.1\times 10^{5}$} & 
$1.3\times 10^{4}$ & 
$1.1\times 10^{4} $\\ 
\hline
%
%
%
%
%
\multirow{2}*{$B^-\to \pi^+ e^-e^-$} &
$(ub)(ud)(ee)$ &
$-$ & 
$5.1\times 10^{2}$ & 
\newb{$2.2\times 10^{3}$} & 
\newb{$2.6\times 10^{4}$} & 
$2.9\times 10^{2}$ & 
$3.6\times 10^{2} $\\ 
 &
$(ud)(ub)(ee)$ &
$-$ & 
$2.2\times 10^{2}$ & 
\newb{$2.7\times 10^{3}$} & 
\newb{$1.1\times 10^{4}$} & 
$1.2\times 10^{2}$ & 
$1.6\times 10^{2} $\\ 
\hline
%
%
%
\multirow{2}*{$B^-\to K^+ e^-e^-$} &
$(ub)(us)(ee)$ &
$-$ & 
$5.6\times 10^{2}$ & 
\newb{$2.2\times 10^{3}$} & 
\newb{$2.8\times 10^{4}$} & 
$3.2\times 10^{2}$ & 
$4.0\times 10^{2} $\\ 
 &
$(us)(ub)(ee)$ &
$-$ & 
$2.2\times 10^{2}$ & 
\newb{$2.7\times 10^{3}$} & 
\newb{$1.1\times 10^{4}$} & 
$1.2\times 10^{2}$ & 
$1.5\times 10^{2} $\\ 
\hline
%
%
%
%
\multirow{4}*{$B^-\to D^+ e^-e^-$} &
$(ub)(cd)(ee)$ &
$-$ & 
$5.4\times 10^{3}$ & 
\newb{$6.8\times 10^{4}$} & 
\newb{$6.8\times 10^{4}$} & 
$3.1\times 10^{3}$ & 
$3.1\times 10^{3} $\\ 
 &
$(cd)(ub)(ee)$ &
$-$ & 
$2.2\times 10^{3}$ & 
\newb{$2.8\times 10^{4}$} & 
\newb{$2.8\times 10^{4}$} & 
$1.3\times 10^{3}$ & 
$1.3\times 10^{3} $\\ 
 &
$(ud)(cb)(ee)$ &
$-$ & 
$-$ & 
\newb{$5.4\times 10^{4}$} & 
\newb{$2.2\times 10^{4}$} & 
\newb{$1.5\times 10^{4}$} & 
\newb{$6.2\times 10^{3}$}\\ 
 &
$(cb)(ud)(ee)$ &
$-$ & 
$-$ & 
\newb{$2.2\times 10^{4}$} & 
\newb{$5.4\times 10^{4}$} & 
\newb{$6.2\times 10^{3}$} & 
\newb{$1.5\times 10^{4}$}\\ 
\hline
\hline
%
%
\multirow{2}*{$K^-\to \pi^+ \mu^-\mu^-$} &
$(us)(ud)(\mu\mu)$ &
$-$ & 
$3.0\times 10^{1}$ & 
\newb{$3.4\times 10^{2}$} & 
\newb{$1.5\times 10^{3}$} & 
$1.7\times 10^{1}$ & 
$2.2\times 10^{1} $\\ 
 &
$(ud)(us)(\mu\mu)$ &
$-$ & 
$3.4\times 10^{1}$ & 
\newb{$3.0\times 10^{2}$} & 
\newb{$1.7\times 10^{3}$} & 
$1.9\times 10^{1}$ & 
$2.4\times 10^{1} $\\ 
\hline
%
%
\multirow{2}*{$D^-\to \pi^+ \mu^-\mu^-$} &
$(cd)(ud)(\mu\mu)$ &
$-$ & 
$2.3\times 10^{3}$ & 
\newb{$1.1\times 10^{5}$} & 
\newb{$2.4\times 10^{4}$} & 
$1.6\times 10^{3}$ & 
$1.3\times 10^{3} $\\ 
 &
$(ud)(cd)(\mu\mu)$ &
$-$ & 
$2.4\times 10^{3}$ & 
\newb{$1.2\times 10^{5}$} & 
\newb{$2.3\times 10^{4}$} & 
$1.7\times 10^{3}$ & 
$1.3\times 10^{3} $\\ 
\hline
%
%
 &
$(cd)(us)(\mu\mu)$ &
$-$ & 
$7.7\times 10^{4}$ & 
\newb{$5.9\times 10^{4}$} & 
\newb{$5.9\times 10^{4}$} & 
\impb{$1.6\times 10^{4}$} & 
\impb{$1.6\times 10^{4}$}\\ 
$D^-\to K^+ \mu^-\mu^-$ &
$(us)(cd)(\mu\mu)$ &
$-$ & 
$7.2\times 10^{4}$ & 
\newb{$5.9\times 10^{4}$} & 
\newb{$5.9\times 10^{4}$} & 
\impb{$1.6\times 10^{4}$} & 
\impb{$1.6\times 10^{4}$}\\ 
$D_s^-\to \pi^+ \mu^-\mu^-$ &
$(cs)(ud)(\mu\mu)$ &
$-$ & 
$5.9\times 10^3$ & 
\newb{$7.3\times 10^{4}$} & 
\newb{$7.3\times 10^{5}$} & 
{$3.3\times 10^{3}$} & 
{$3.3\times 10^{3}$}\\ 
 &
$(ud)(cs)(\mu\mu)$ &
$-$ & 
$5.9\times 10^3$ & 
\newb{$7.4\times 10^{4}$} & 
\newb{$7.4\times 10^{4}$} & 
{$3.3\times 10^{3}$} & 
{$3.3\times 10^{3}$}\\ 
\hline
%
%
%
\multirow{2}*{$D_s^-\to K^+ \mu^-\mu^-$} &
$(cs)(us)(\mu\mu)$ &
$-$ & 
$4.0\times 10^{3}$ & 
\newb{$2.0\times 10^{5}$} & 
\newb{$3.6\times 10^{4}$} & 
$2.8\times 10^{3}$ & 
$2.3\times 10^{3} $\\ 
 &
$(us)(cs)(\mu\mu)$ &
$-$ & 
$3.6\times 10^{3}$ & 
\newb{$1.8\times 10^{5}$} & 
\newb{$4.0\times 10^{4}$} & 
$2.6\times 10^{3}$ & 
$2.0\times 10^{3} $\\ 
\hline
%
%
\multirow{2}*{$B^-\to \pi^+ \mu^-\mu^-$} &
$(ub)(ud)(\mu\mu)$ &
$-$ & 
$2.1\times 10^{2}$ & 
\newb{$9.2\times 10^{2}$} & 
\newb{$1.1\times 10^{4}$} & 
$1.2\times 10^{2}$ & 
$1.5\times 10^{2} $\\ 
 &
$(ud)(ub)(\mu\mu)$ &
$-$ & 
$9.2\times 10^{1}$ & 
\newb{$1.1\times 10^{3}$} & 
\newb{$4.6\times 10^{3}$} & 
$5.2\times 10^{1}$ & 
$6.5\times 10^{1} $\\ 
\hline
%
%
%
%
\multirow{2}*{$B^-\to K^+ \mu^-\mu^-$} &
$(ub)(us)(\mu\mu)$ &
$-$ & 
$6.6\times 10^{2}$ & 
\newb{$2.5\times 10^{3}$} & 
\newb{$3.3\times 10^{4}$} & 
$3.7\times 10^{2}$ & 
$4.7\times 10^{2} $\\ 
 &
$(us)(ub)(\mu\mu)$ &
$-$ & 
$2.5\times 10^{2}$ & 
\newb{$3.2\times 10^{3}$} & 
\newb{$1.3\times 10^{4}$} & 
$1.4\times 10^{2}$ & 
$1.8\times 10^{2} $\\ 
\hline
%
%
%
%
\multirow{4}*{$B^-\to D^+ \mu^-\mu^-$} &
$(ub)(cd)(\mu\mu)$ &
$-$ & 
$2.8\times 10^{3}$ & 
\newb{$3.5\times 10^{4}$} & 
\newb{$3.5\times 10^{4}$} & 
$1.6\times 10^{3}$ & 
$1.6\times 10^{3} $\\ 
 &
$(cd)(ub)(\mu\mu)$ &
$-$ & 
$1.2\times 10^{3}$ & 
\newb{$1.5\times 10^{4}$} & 
\newb{$1.5\times 10^{4}$} & 
$6.6\times 10^{2}$ & 
$6.6\times 10^{2} $\\ 
 &
$(ud)(cb)(\mu\mu)$ &
$-$ & 
$-$ & 
\newb{$2.8\times 10^{4}$} & 
\newb{$1.2\times 10^{4}$} & 
\newb{$7.8\times 10^{3}$} & 
\newb{$3.2\times 10^{3}$}\\ 
 &
$(cb)(ud)(\mu\mu)$ &
$-$ & 
$-$ & 
\newb{$1.2\times 10^{4}$} & 
\newb{$2.8\times 10^{4}$} & 
\newb{$3.2\times 10^{3}$} & 
\newb{$7.8\times 10^{3}$}\\ 
\hline
%
%
%
\multirow{4}*{$B^-\to D_s^+ \mu^-\mu^-$} &
$(ub)(cs)(\mu\mu)$ &
$-$ & 
$2.2\times 10^{3}$ & 
\newb{$2.8\times 10^{4}$} & 
\newb{$2.8\times 10^{4}$} & 
$1.3\times 10^{3}$ & 
$1.3\times 10^{3} $\\ 
 &
$(cs)(ub)(\mu\mu)$ &
$-$ & 
$9.5\times 10^{2}$ & 
\newb{$1.2\times 10^{4}$} & 
\newb{$1.2\times 10^{4}$} & 
$5.4\times 10^{2}$ & 
$5.4\times 10^{2} $\\ 
 &
$(us)(cb)(\mu\mu)$ &
$-$ & 
$-$ & 
\newb{$2.2\times 10^{4}$} & 
\newb{$9.5\times 10^{3}$} & 
\newb{$6.2\times 10^{3}$} & 
\newb{$2.6\times 10^{3}$}\\ 
 &
$(cb)(us)(\mu\mu)$ &
$-$ & 
$-$ & 
\newb{$9.5\times 10^{3}$} & 
\newb{$2.2\times 10^{4}$} & 
\newb{$2.6\times 10^{3}$} & 
\newb{$6.2\times 10^{3}$}\\ 
\hline
\hline
%
%
\multirow{2}*{$K^-\to \pi^+ e^-\mu^-$} &
$(us)(ud)(e\mu)$ &
$-$ & 
$1.4\times 10^{1}$ & 
\newb{$1.6\times 10^{2}$} & 
\newb{$6.9\times 10^{2}$} & 
$7.8$ & 
$9.8$\\ 
 &
$(ud)(us)(e\mu)$ &
$-$ & 
$1.6\times 10^{1}$ & 
\newb{$1.4\times 10^{2}$} & 
\newb{$7.8\times 10^{2}$} & 
$8.8$ & 
$1.1\times 10^1$\\ 
\hline
%
%
\multirow{2}*{$D^-\to \pi^+ e^-\mu^-$} &
$(cd)(ud)(e\mu)$ &
$-$ & 
$4.8\times 10^{3}$ & 
\newb{$2.4\times 10^{5}$} & 
\newb{$5.0\times 10^{4}$} & 
$3.4\times 10^{3}$ & 
$2.7\times 10^{3} $\\ 
 &
$(ud)(cd)(e\mu)$ &
$-$ & 
$5.0\times 10^{3}$ & 
\newb{$2.5\times 10^{5}$} & 
\newb{$4.8\times 10^{4}$} & 
$3.5\times 10^{3}$ & 
$2.8\times 10^{3} $\\ 
\hline
%
%
 &
$(cd)(us)(e\mu)$ &
$-$ & 
$2.3\times 10^{4}$ & 
\newb{$1.1\times 10^{5}$} & 
\newb{$1.1\times 10^{5}$} & 
$1.3\times 10^{4}$ & 
$1.3\times 10^{4} $\\ 
$D^-\to K^+ e^-\mu^-$ &
$(us)(cd)(e\mu)$ &
$-$ & 
$2.2\times 10^{4}$ & 
\newb{$1.1\times 10^{5}$} & 
\newb{$1.1\times 10^{5}$} & 
$1.2\times 10^{4}$ & 
$1.2\times 10^{4} $\\ 
$D_s^-\to \pi^+ e^-\mu^-$ &
$(cs)(ud)(e\mu)$ &
$-$ & 
$1.1\times 10^4$ & 
\newb{$1.4\times 10^{5}$} & 
\newb{$1.4\times 10^{5}$} & 
{$6.2\times 10^{3}$} & 
{$6.2\times 10^{3}$}\\ 
 &
$(ud)(cs)(e\mu)$ &
$-$ & 
$1.1\times 10^4$ & 
\newb{$1.4\times 10^{5}$} & 
\newb{$1.4\times 10^{5}$} & 
{$6.3\times 10^{3}$} & 
{$6.3\times 10^{3}$}\\ 
\hline
%
%
%
\multirow{2}*{$D_s^-\to K^+ e^-\mu^-$} &
$(cs)(us)(e\mu)$ &
$-$ & 
$8.8\times 10^{3}$ & 
\newb{$4.4\times 10^{5}$} & 
\newb{$7.9\times 10^{4}$} & 
$6.2\times 10^{3}$ & 
$5.0\times 10^{3} $\\ 
 &
$(us)(cs)(e\mu)$ &
$-$ & 
$7.9\times 10^{3}$ & 
\newb{$3.9\times 10^{5}$} & 
\newb{$8.8\times 10^{4}$} & 
$5.6\times 10^{3}$ & 
$4.4\times 10^{3} $\\ 
\hline
%
%
%
\multirow{2}*{$B^-\to \pi^+ e^-\mu^-$} &
$(ub)(ud)(e\mu)$ &
$-$ & 
$9.3\times 10^{2}$ & 
\newb{$4.0\times 10^{3}$} & 
\newb{$4.6\times 10^{4}$} & 
$5.2\times 10^{2}$ & 
$6.6\times 10^{2} $\\ 
 &
$(ud)(ub)(e\mu)$ &
$-$ & 
$4.0\times 10^{2}$ & 
\newb{$4.9\times 10^{3}$} & 
\newb{$2.0\times 10^{4}$} & 
$2.2\times 10^{2}$ & 
$2.8\times 10^{2} $\\ 
\hline
%
%
%
\multirow{2}*{$B^-\to K^+ e^-\mu^-$} &
$(ub)(us)(e\mu)$ &
$-$ & 
$9.2\times 10^{2}$ & 
\newb{$3.5\times 10^{3}$} & 
\newb{$4.6\times 10^{4}$} & 
$5.2\times 10^{2}$ & 
$6.5\times 10^{2} $\\ 
 &
$(us)(ub)(e\mu)$ &
$-$ & 
$3.5\times 10^{2}$ & 
\newb{$4.4\times 10^{3}$} & 
\newb{$1.8\times 10^{4}$} & 
$2.0\times 10^{2}$ & 
$2.5\times 10^{2} $\\ 
\hline
%
%
%
\multirow{4}*{$B^-\to D^+ e^-\mu^-$} &
$(ub)(cd)(e\mu)$ &
$-$ & 
$3.2\times 10^{3}$ & 
\newb{$4.0\times 10^{4}$} & 
\newb{$4.0\times 10^{4}$} & 
$1.8\times 10^{3}$ & 
$1.8\times 10^{3} $\\ 
 &
$(cd)(ub)(e\mu)$ &
$-$ & 
$1.3\times 10^{3}$ & 
\newb{$1.7\times 10^{4}$} & 
\newb{$1.7\times 10^{4}$} & 
$7.5\times 10^{2}$ & 
$7.5\times 10^{2} $\\ 
 &
$(ud)(cb)(e\mu)$ &
$-$ & 
$-$ & 
\newb{$3.2\times 10^{4}$} & 
\newb{$1.3\times 10^{4}$} & 
\newb{$8.9\times 10^{3}$} & 
\newb{$3.7\times 10^{3}$}\\ 
 &
$(cb)(ud)(e\mu)$ &
$-$ & 
$-$ & 
\newb{$1.3\times 10^{4}$} & 
\newb{$3.2\times 10^{4}$} & 
\newb{$3.7\times 10^{3}$} & 
\newb{$8.9\times 10^{3}$}\\ 
\hline
\hline
\end{NiceTabular}
\caption{
\scriptsize
Upper bounds on the Wilson coefficients $C_{4(ijkl)}^{XYZ(\ell_1\ell_2)}$ and $C_{5(ijkl)}^{XYZ(\ell_1\ell2)}$ extracted from the experimental limits on the LNV meson decays $M_1\to M_2\, \ell_1^- \ell_2^-$ shown in Table (\ref{tab:expbounds_mesondecays}), with and without QCD corrections.
The bounds in gray background are new results present only after including the effects of QCD corrections, which is the case for all the bounds on $C_{4(ijkl)}^{XYZ(\ell_1\ell_2)}$.
Note that the operators bounded by the processes $D^-\to K^+ \ell_1^-\ell_2^-$ and $D_s^-\to \pi^+ \ell_1^-\ell_2^-$ mix under RG evolution, so these processes put bounds on the same sets of operators. The operators in light-gray background are improved due to this mixing.
\normalsize
}
\label{tab:limits_mesons_O45_ee_noCKMbasis}
\normalsize
\end{table}
\FloatBarrier
%
%

\begin{table}[h!]\footnotesize
\centering
\begin{NiceTabular}{c|c|c|cccc}[colortbl-like]
\multicolumn{2}{c}{ }  &
\multicolumn{1}{c}{Tree Level} & \multicolumn{3}{c}{With QCD Corrections}\\
\hline
Process &
$(ij)(kl)(\ell_1\ell_2)$ & 
$\left|C_{6(ijkl)}^{XYZ}\right|$ &
$\left|C_{6(ijkl)}^{XXX}\right|$ & $\left|C_{6(ijkl)}^{LLR,RRL}\right|$ & $\left|C_{6(ijkl)}^{LRL,RLR}\right|$ & $\left|C_{6(ijkl)}^{LRR,RLL}\right|$
\\
\hline
\hline
$K^-\to \pi^+ e^-\mu^-$ &
$(us)(ud)(e\mu)$ &
$2.6\times 10^{2}$ &
$2.8\times 10^{2}$ & 
$2.1\times 10^{2}$ & 
$2.7\times 10^{2}$ & 
$2.7\times 10^{2}$ \\ 
\hline
%
%
%
$D^-\to \pi^+ e^-\mu^-$ &
$(cd)(ud)(e\mu)$ &
$2.3\times 10^{4}$ &
$1.9\times 10^{4}$ & 
$2.5\times 10^{4}$ & 
$2.4\times 10^{4}$ & 
$2.4\times 10^{4}$ \\ 
\hline
%
%
%
$D^-\to K^+ e^-\mu^-$ &
$(cd)(us)(e\mu)$ &
$1.2\times 10^{5}$ &
$1.1\times 10^{5}$ & 
$1.1\times 10^{5}$ & 
$1.2\times 10^{5}$ & 
$1.2\times 10^{5}$ \\ 
%
$D_s^-\to \pi^+ e^-\mu^-$ &
$(cs)(ud)(e\mu)$ &
$5.1\times 10^4$ &
$4.7\times 10^{4}$ & 
$4.7\times 10^{4}$ & 
$5.3\times 10^{4}$ & 
$5.3\times 10^{4}$ \\ 
\hline
%
%
%
$D_s^-\to K^+ e^-\mu^-$ &
$(cs)(us)(e\mu)$ &
$4.1\times 10^{4}$ &
$3.4\times 10^{4}$ & 
$4.5\times 10^{4}$ & 
$4.4\times 10^{4}$ & 
$4.4\times 10^{4}$ \\ 
\hline
$B^-\to \pi^+ e^-\mu^-$ &
$(ub)(ud)(e\mu)$ &
$1.6\times 10^{3}$ &
$1.7\times 10^{3}$ & 
$1.3\times 10^{3}$ & 
$1.7\times 10^{3}$ & 
$1.7\times 10^{3}$ \\ 
\hline
$B^-\to K^+ e^-\mu^-$ &
$(ub)(us)(e\mu)$ &
$1.4\times 10^{3}$ &
$1.6\times 10^{3}$ & 
$1.2\times 10^{3}$ & 
$1.5\times 10^{3}$ & 
$1.5\times 10^{3}$ \\ 
\hline
%
%
%
\multirow{2}*{$B^-\to D^+ e^-\mu^-$} &
$(ub)(cd)(e\mu)$ &
$7.0\times 10^{3}$ &
$6.5\times 10^{3}$ & 
$6.5\times 10^{3}$ & 
$7.4\times 10^{3}$ & 
$7.4\times 10^{3}$ \\ 
%
 &
$(ud)(cb)(e\mu)$ &
$-$ &
\newb{$3.7\times 10^{4}$} & 
\newb{$3.7\times 10^{4}$} & 
$-$ & 
$-$ \\ 
\hline
\hline
\end{NiceTabular}
\caption{Upper bounds on the Wilson coefficient $C_{6(ijkl)}^{XYZ(e\mu)}$ extracted from the experimental limits on the LNV meson decays $M_1^-\to M_2^-\, e^- \mu^-$ shown in Table (\ref{tab:expbounds_mesondecays}), with and without QCD corrections.
The constrained operators involve different lepton flavors ($\ell_1\neq \ell_2$) since the operator $\mathcal O_{6(ijkl)}^{XYZ(\ell_1\ell_2)}$ vanishes otherwise.
The bounds in gray background are new results present only after including the effects of QCD corrections.
Note that the operators bounded by the processes $D^-\to K^+ \ell_1^-\ell_2^-$ and $D_s^-\to \pi^+ \ell_1^-\ell_2^-$ mix under RG evolution, so these processes put bounds on the same sets of operators.
The limits for $C_{6(ijkl)}^{LRL,RLR}$ and $C_{6(ijkl)}^{LRR,RLL}$ are the same due to the relation $\mathcal O_{6(ijkl)}^{LRZ}=-\mathcal O_{6(klij)}^{RLZ}$.
}
\label{tab:limits_mesons_6_emu_noCKMbasis}
\end{table}

\begin{table}[h!]\footnotesize
\centering
\begin{NiceTabular}{c|c|c|c}[colortbl-like]
\multicolumn{2}{c}{ }  &
\multicolumn{1}{c}{Tree Level} & {With QCD Corrections}\\
\hline
Process &
$(ij)(kl)(\ell_1\ell_2)$ & 
$\left|C_{7(ijkl)}^{XZZ}\right|$ &
$\left|C_{7(ijkl)}^{LRR,RLL}\right|$\\
\hline
\hline
\multirow{2}*{$K^-\to \pi^+ e^-\mu^-$} &
$(us)(ud)(e\mu)$ &
$-$ &
\newb{$3.2\times 10^{3}$}\\ 
& $(ud)(us)(e\mu)$& $-$ & \newb{$3.2\times 10^3$}\\
\hline
%
%
%
\multirow{2}*{$D^-\to \pi^+ e^-\mu^-$} &
$(cd)(ud)(e\mu)$ &
$-$ &
\newb{$2.9\times 10^{5}$} 
\\
 & $(ud)(cd)(e\mu)$ & $-$ & \newb{$2.9\times 10^5$}\\
\hline
%
%
%
\multirow{2}*{$D^-\to K^+ e^-\mu^-$} &
$(cs)(ud)(e\mu)$ &
$-$ &
\newb{$1.5\times 10^6$} \\
& $(ud)(cs)(e\mu)$ & $-$ & \newb{$1.5\times 10^6$}\\
\hline
\multirow{2}*{$D_s^-\to \pi^+ e^-\mu^-$} &
$(cd)(us)(e\mu)$ &
$-$ &
\newb{$6.3\times 10^{5}$} \\
 & $(us)(cd)(e\mu)$ & $-$ & \newb{$6.3\times 10^{5}$}\\
\hline
%
%
%
\multirow{2}*{$D_s^-\to K^+ e^-\mu^-$} &
$(cs)(us)(e\mu)$ &
$-$ &
\newb{$5.2\times 10^{5}$}\\
& $(us)(cs)(e\mu)$ & $-$ & \newb{$5.2\times 10^{5}$}\\
\hline
\multirow{2}*{$B^-\to \pi^+ e^-\mu^-$} &
$(ub)(ud)(e\mu)$ &
$-$ &
\newb{$2.0\times 10^{4}$} \\ 
 & $(ud)(ub)(e\mu)$ & $-$ & \newb{$2.0\times 10^{4}$}\\
\hline
\multirow{2}*{$B^-\to K^+ e^-\mu^-$} &
$(ub)(us)(e\mu)$ &
$-$ &
\newb{$1.8\times 10^{4}$} \\ 
 & $(us)(ub)(e\mu)$ & $-$ & \newb{$1.8\times 10^{4}$}\\
\hline
%
%
%
\multirow{2}*{$B^-\to D^+ e^-\mu^-$} &
$(cb)(ud)(e\mu)$ &
$-$ &
\newb{$8.8\times 10^{4}$} 
 \\
& $(ud)(cb)(e\mu)$ & $-$ & \newb{$8.8\times 10^{4}$}\\
\hline
\hline
\end{NiceTabular}
\caption{Upper bounds on the Wilson coefficient $C_{7(ijkl)}^{XZZ(e\mu)}$ extracted from the experimental limits on the LNV meson decays $M_1^-\to M_2^-\, e^- \mu^-$ shown in Table (\ref{tab:expbounds_mesondecays}), with and without QCD corrections.
All the bounds shown in this table (in
gray) are new results present only after including the effects of QCD corrections.
The constrained operators involve different lepton flavors ($\ell_1\neq \ell_2$) since the operator $\mathcal O_{7(ijkl)}^{XZZ(\ell_1\ell_2)}$ vanishes otherwise.
Note that these bounds come from the mixing with the operator $\mathcal O_{6(ijkl)}^{XYZ(e\mu)}$, which is constrained at the tree level [see Table\,(\ref{tab:limits_mesons_6_emu_noCKMbasis})].
In particular, the second bound shown for each process is due to the relation $\mathcal O_{6(ijkl)}^{RLR,LRL}=-\mathcal O_{6(klij)}^{LRR,RLL}$ and the mixing between $\mathcal O_{6(ijkl)}^{LRR,RLL}$ and $\mathcal O_{7(ijkl)}^{LRR,RLL}$ [see appendix \ref{app:umatrices}].
}
\label{tab:limits_mesons_7_emu_noCKMbasis}
\end{table}

\begin{table}[h!]\footnotesize
\centering
\begin{NiceTabular}{c|c|cc|ccc}[colortbl-like]
\multicolumn{2}{c}{ }  & \multicolumn{2}{c}{Tree Level} & \multicolumn{3}{c}{With QCD Corrections}\\
\hline
Process &$(ij)(kl)(\ell_1\ell_2)$ & 
$\left|C_{1(ijkl)}^{XYZ}\right|$ & $\left|C_{2(ijkl)}^{XXX}\right|$ & $\left|C_{1(ijkl)}^{XXZ}\right|$ & $\left|C_{1(ijkl)}^{(LR,RL)Z}\right|$ & $\left|C_{2(ijkl)}^{XXX}\right|$ \\
\hline
\hline
$\tau^-\to e^+ \pi^- \pi^-$ & $(ud)(ud)(e\tau)$ &
$2.9\times 10^{3}$ & $-$ &
{$1.5\times 10^{3}$} & 
\impb{$9.5\times 10^{2}$} & 
\newb{$2.9\times 10^{5}$} \\ 
\hline
$\tau^-\to e^+ \pi^- K^-$ & $(ud)(us)(e\tau)$ &
$3.3\times 10^3$ & $-$ & 
{$1.7\times 10^{3}$} & 
\impb{$1.1\times 10^{3}$} & 
\newb{$3.3\times 10^{5}$} \\ 
\hline
$\tau^-\to e^+ K^- K^-$ & $(us)(us)(e\tau)$ &
$7.1\times 10^3$ & $-$ &
{$3.6\times 10^{3}$} & 
\impb{$2.3\times 10^{3}$} & 
\newb{$7.1\times 10^{5}$} \\ 
\hline
\hline
$\tau^-\to \mu^+ \pi^- \pi^-$ & $(ud)(ud)(\mu\tau)$ &
$4.0\times 10^{3}$ & $-$ &
\impb{$2.0\times 10^{3}$} & 
\impb{$1.3\times 10^{3}$} & 
\newb{$4.0\times 10^{5}$} \\ 
\hline
$\tau^-\to \mu^+ \pi^- K^-$ & $(ud)(us)(\mu\tau)$ &
$4.1\times 10^3$ & $-$ &
{$2.1\times 10^{3}$} & 
\impb{$1.4\times 10^{3}$} & 
\newb{$4.1\times 10^{5}$} \\ 
\hline
$\tau^-\to \mu^+ K^- K^-$ & $(us)(us)(\mu\tau)$ &
$8.6\times 10^3$ & $-$ & 
\impb{$4.3\times 10^{3}$} & 
\impb{$2.8\times 10^{3}$} & 
\newb{$8.6\times 10^{5}$} \\ 
\hline
\hline
\end{NiceTabular}
\caption{Upper bounds on the Wilson coefficients $C_{1(ijkl)}^{XYZ(\ell\tau)}$ and  $C_{2(ijkl)}^{XXX(\ell\tau)}$ extracted from the experimental limits on the LNV tau decays $\tau^-\to \ell^+ M_1^- M_2^-$ shown in Table (\ref{tab:expbounds_taudecays}), with and without QCD corrections.
The bounds in gray background are new results present only after including the effects of QCD corrections.
The bounds which improve by at least a factor of 2 with respect to the corresponding tree-level bound are shown in light-gray.}\label{tab:limits_taus_O12}
\end{table}
\begin{table}[h!]\footnotesize
\centering
\begin{NiceTabular}{c|c|c|cc}[colortbl-like]
\multicolumn{2}{c}{ }  & \multicolumn{1}{c}{Tree Level} & \multicolumn{2}{c}{With QCD Corrections}\\
\hline
Process &$(ij)(kl)(\ell_1\ell_2)$ & 
$\left|C_{3(ijkl)}^{XYZ}\right|$ & $\left|C_{3(ijkl)}^{XXZ}\right|$ & $\left|C_{3(ijkl)}^{(LR,RL)Z}\right|$
\\
\hline
\hline
$\tau^-\to e^+ \pi^- \pi^-$ & $(ud)(ud)(e\tau)$ &
$4.5\times 10^{4}$ &
$5.9\times 10^{4}$ & 
$5.2\times 10^4$\\ 
\hline
$\tau^-\to e^+ \pi^- K^-$ & $(ud)(us)(e\tau)$ &
$4.4\times 10^4$ &
$5.8\times 10^{4}$ & 
$5.1\times 10^{4}$ \\ 
\hline
$\tau^-\to e^+ K^- K^-$ & $(us)(us)(e\tau)$ &
$7.4\times 10^4$ &
$9.8\times 10^{4}$ & 
$8.5\times 10^4$ \\ 
\hline
\hline
$\tau^-\to \mu^+ \pi^- \pi^-$ & $(ud)(ud)(\mu\tau)$ &
$6.4\times 10^{4}$ &
$8.5\times 10^{4}$ & 
$7.4\times 10^4$\\ 
\hline
$\tau^-\to \mu^+ \pi^- K^-$ & $(ud)(us)(\mu\tau)$ &
$5.6\times 10^4$ &
$7.3\times 10^{4}$ & 
$6.4\times 10^{4}$\\ 
\hline
$\tau^-\to \mu^+ K^- K^-$ & $(us)(us)(\mu\tau)$ &
$9.3\times 10^4$ &
$1.2\times 10^{5}$ & 
$1.1\times 10^5$\\ 
\hline
\hline
\end{NiceTabular}
\caption{Upper bounds on the Wilson coefficient $C_{3(ijkl)}^{XYZ(\ell\tau)}$ extracted from the experimental limits on the LNV tau decays $\tau^-\to \ell^+ M_1^- M_2^-$ shown in Table (\ref{tab:expbounds_taudecays}), with and without QCD corrections.
For these operators, existing tree-level bounds are not improved by QCD effects.}
\label{tab:limits_taus_O3}
\end{table}

\begin{table}[h!]\scriptsize
\centering
\begin{NiceTabular}{c|c|cc|cccc}[colortbl-like]
\multicolumn{2}{c}{ }  & \multicolumn{2}{c}{Tree Level} & \multicolumn{4}{c}{With QCD Corrections}\\
\hline
Process &$(ij)(kl)(\ell_1\ell_2)$ & $\left|C_{4(ijkl)}^{XYZ}\right|$ & $\left|C_{5(ijkl)}^{XYZ}\right|$ & $\left|C_{4(ijkl)}^{XXZ}\right|$ & $\left|C_{4(ijkl)}^{(LR,RL)Z}\right|$ & $\left|C_{5(ijkl)}^{XXZ}\right|$ & $\left|C_{5(ijkl)}^{(LR,RL)Z}\right|$
\\
\hline
\hline
%
$\tau^-\to e^+ \pi^- \pi^-$ &
$(ud)(ud)(e\tau)$ &
$-$ & 
$7.7\times 10^{3}$ & 
\newb{$2.6\times 10^{5}$} & 
\newb{$7.7\times 10^{5}$} & 
$6.1\times 10^{3}$ & 
$4.9\times 10^{3} $\\ 
\hline
%
\multirow{2}*{$\tau^- \to e^+ \pi^- K^-$} &
$(ud)(us)(e\tau)$ &
$-$ & 
$7.0\times 10^{3}$ & 
\newb{$7.8\times 10^{4}$} & 
\newb{$3.5\times 10^{5}$} & 
$4.0\times 10^{3}$ & 
$5.0\times 10^{3} $\\ 
 &
$(us)(ud)(e\tau)$ &
$-$ & 
$7.8\times 10^{3}$ & 
\newb{$7.0\times 10^{4}$} & 
\newb{$3.9\times 10^{5}$} & 
$4.4\times 10^{3}$ & 
$5.6\times 10^{3} $\\ 
\hline
%
%
%
$\tau^-\to e^+ K^- K^-$ &
$(us)(us)(e\tau)$ &
$-$ & 
$1.8\times 10^{4}$ & 
\newb{$6.0\times 10^{5}$} & 
\newb{$1.8\times 10^{6}$} & 
$1.4\times 10^{4}$ & 
$1.1\times 10^{4} $\\ 
\hline
\hline
%
%
$\tau^-\to \mu^+ \pi^- \pi^-$ &
$(ud)(ud)(\mu\tau)$ &
$-$ & 
$1.1\times 10^{4}$ & 
\newb{$3.6\times 10^{5}$} & 
\newb{$1.1\times 10^{6}$} & 
$8.6\times 10^{3}$ & 
$6.9\times 10^{3} $\\ 
\hline
%
\multirow{2}*{$\tau^- \to \mu^+ \pi^- K^-$} &
$(ud)(us)(\mu\tau)$ &
$-$ & 
$8.7\times 10^{3}$ & 
\newb{$9.8\times 10^{4}$} & 
\newb{$4.4\times 10^{5}$} & 
$4.9\times 10^{3}$ & 
$6.2\times 10^{3} $\\ 
 &
$(us)(ud)(\mu\tau)$ &
$-$ & 
$9.8\times 10^{3}$ & 
\newb{$8.7\times 10^{4}$} & 
\newb{$4.9\times 10^{5}$} & 
$5.5\times 10^{3}$ & 
$6.9\times 10^{3} $\\ 
\hline
%
%
%
$\tau^-\to \mu^+ K^- K^-$ &
$(us)(us)(\mu\tau)$ &
$-$ & 
$2.2\times 10^{4}$ & 
\newb{$7.4\times 10^{5}$} & 
\newb{$2.2\times 10^{6}$} & 
$1.8\times 10^{4}$ & 
$1.4\times 10^{4} $\\ 
\hline
\hline
\end{NiceTabular}
\caption{Upper bounds on the Wilson coefficients $C_{4(ijkl)}^{XYZ(\ell\tau)}$ and $C_{5(ijkl)}^{XYZ(\ell\tau)}$ extracted from the experimental limits on the LNV tau decays $\tau^-\to \ell^+ M_1^- M_2^-$ shown in Table (\ref{tab:expbounds_taudecays}), with and without QCD corrections.
The bounds in gray background are new results present only after including the effects of QCD corrections, which is the case for all the bounds on $C_{4(ijkl)}^{XYZ(\ell\tau)}$.}
\label{tab:limits_taus_O45}
\normalsize
\end{table}
%
%
%
\FloatBarrier
\begin{table}[h!]\footnotesize
\centering
\begin{NiceTabular}{c|c|c|cccc}[colortbl-like]
\multicolumn{2}{c}{ }  &
\multicolumn{1}{c}{Tree Level} & \multicolumn{4}{c}{With QCD Corrections}\\
\hline
Process &
$(ij)(kl)(\ell_1\ell_2)$ & 
$\left|C_{6(ijkl)}^{XYZ}\right|$ &
$\left|C_{6(ijkl)}^{XXX}\right|$ & $\left|C_{6(ijkl)}^{LLR,RRL}\right|$ & $\left|C_{6(ijkl)}^{LRL,RLR}\right|$ & $\left|C_{6(ijkl)}^{LRR,RLL}\right|$ \\
\hline
\hline
$\tau^-\to e^+ \pi^- \pi^-$ &
$(ud)(ud)(e\tau)$ &
$1.3\times 10^{4}$ &
$-$ & 
$-$ & 
$1.4\times 10^{4}$ & 
$1.4\times 10^{4}$ \\ 
\hline
%
%
$\tau^-\to e^+ \pi^- K^-$ &
$(ud)(us)(e\tau)$ &
$1.5\times 10^{4}$ &
$1.6\times 10^4$ & 
$1.2\times 10^{4}$ & 
$1.6\times 10^{4}$ & 
$1.6\times 10^{4}$ \\ 
\hline
%
%
$\tau^-\to e^+ K^- K^-$ &
$(us)(us)(e\tau)$ &
$2.8\times 10^{4}$ &
$-$ & 
$-$ & 
$2.9\times 10^{4}$ & 
$2.9\times 10^{4}$ \\ 
\hline
\hline
%
%
$\tau^-\to \mu^+ \pi^- \pi^-$ &
$(ud)(ud)(\mu\tau)$ &
$1.9\times 10^{4}$ &
$-$ & 
$-$ & 
$2.0\times 10^{4}$ & 
$2.0\times 10^{4}$ \\ 
\hline
%
%
$\tau^-\to \mu^+ \pi^- K^-$ &
$(ud)(us)(\mu\tau)$ &
$1.9\times 10^{4}$ &
$2.1\times 10^4$ & 
$1.6\times 10^{4}$ & 
$2.0\times 10^{4}$ & 
$2.0\times 10^{4}$ \\ 
\hline
%
%
$\tau^-\to \mu^+ K^- K^-$ &
$(us)(us)(\mu\tau)$ &
$3.5\times 10^{4}$ &
$-$ & 
$-$ & 
$3.7\times 10^{4}$ & 
$3.7\times 10^{4}$ \\ 
\hline
\hline
\end{NiceTabular}
\caption{Upper bounds on the Wilson coefficient $C_{6(ijkl)}^{XYZ(\ell\tau)}$ extracted from the experimental limits on the LNV tau decays $\tau^-\to \ell^+ M_1^- M_2^-$ shown in Table (\ref{tab:expbounds_taudecays}), with and without QCD corrections.
The operators not showing a bound vanish due to their symmetry. (One can check that $\mathcal O_{6(ijkl)}^{XYZ} = 0$ for $X=Y$, $i=k$, and $j=l$.)
}
\label{tab:limits_taus_6_noCKMbasis}
\end{table}
%
%
%
\begin{table}[h!]\footnotesize
\centering
\begin{NiceTabular}{c|c|c|c}[colortbl-like]
\multicolumn{2}{c}{ }  &
\multicolumn{1}{c}{Tree Level} & {With QCD Corrections}\\
\hline
Process &
$(ij)(kl)(\ell_1\ell_2)$ & 
$\left|C_{7(ijkl)}^{XZZ}\right|$ &
$\left|C_{7(ijkl)}^{LRR,RLL}\right|$ \\
\hline
\hline
$\tau^-\to e^+ \pi^- \pi^-$ &
$(ud)(ud)(e\tau)$ &
$-$ &
\newb{$1.7\times 10^{5}$} \\ 
\hline
%
%
\multirow{2}*{$\tau^-\to e^+ \pi^- K^-$} &
$(ud)(us)(e\tau)$ &
$-$ &
\newb{$1.9\times 10^{5}$} \\ 
 & $(us)(ud)(e\tau)$ & $-$ & \newb{$1.9\times 10^5$}\\
\hline
%
%
$\tau^-\to e^+ K^- K^-$ &
$(us)(us)(e\tau)$ &
$-$ &
\newb{$3.5\times 10^{5}$} \\ 
\hline
\hline
%
%
$\tau^-\to \mu^+ \pi^- \pi^-$ &
$(ud)(ud)(\mu\tau)$ &
$-$ &
\newb{$2.4\times 10^{5}$} \\ 
\hline
%
%
\multirow{2}*{$\tau^-\to \mu^+ \pi^- K^-$} &
$(ud)(us)(\mu\tau)$ &
$-$ &
\newb{$2.4\times 10^{5}$} \\ 
 & $(us)(ud)(\mu\tau)$ & $-$ & \newb{$2.4\times 10^5$}\\
\hline
%
%
$\tau^-\to \mu^+ K^- K^-$ &
$(us)(us)(\mu\tau)$ &
$-$ &
\newb{$4.4\times 10^{5}$} \\ 
\hline
\hline
\end{NiceTabular}
\caption{Upper bounds on the Wilson coefficient $C_{7(ijkl)}^{XZZ(\ell\tau)}$ extracted from the experimental limits on the LNV tau decays $\tau^-\to \ell^+ M_1^- M_2^-$ shown in Table (\ref{tab:expbounds_taudecays}), with and without QCD corrections.
All the bounds shown in this table (in
gray) are new results present only after including the effects of QCD corrections.
Note that these bounds come from mixing with the operator $\mathcal O_{6(ijkl)}^{XYZ(\ell\tau)}$, which is constrained at the tree level [see Table\,(\ref{tab:limits_taus_6_noCKMbasis})].
}
\label{tab:limits_taus_7_noCKMbasis}
\end{table}
\FloatBarrier

\newpage
\newpage

\newpage
\subsection*{Acknowledgements}\label{sec:acknowledgements}

We thank Nestor Quintero for valuable discussions.
We acknowledge support from Centro de F\'isica Te\'orica de Valpara\'iso (CeFiTeV), project PFE UVA22991/PUENTE,
ANID (Chile) FONDECYT Iniciaci\'on Grant No. 11230879, and ANID REC Convocatoria Nacional Subvenci\'on a Instalaci\'on en la Academia Convocatoria A\~no 2020, PAI77200092.

\begin{appendices}

\section{$\mu$-evolution matrices}\label{app:umatrices}

In this section we list the evolution matrices $\hat U(\mu,\Lambda)$ defined in Eq.~(\ref{eq:RGE-Sol}) and calculated from Eq.\,(\ref{eq:UmuLambda}) with the anomalous dimension matrices obtained in Sec.\,(\ref{sec:QCDcorrections}).
In the calculation we set $\mu=1\,\mbox{GeV}$ and $\Lambda=m_Z$.
These evolution matrices are used in Sec.\,(\ref{sec:RGEinducedbounds}) to find the RGE-improved bounds on the Wilson coefficients.
We organize them in four groups depending on the quark flavor structure of the Wilson coefficients $\mathcal O_{n(ijkl)}$ ($ijkl$ indices): (i) Repeated up-flavor quarks, (ii) Repeated down-flavor quarks, (iii) Repeated up- and down-flavor quarks, (iv) all quark flavors different.
Similarly to the notation used for the anomalous dimensions in Sec.\,(\ref{sec:QCDcorrections}), we assume in the matrices below that the flavor indices $i,j,k,l$ are different, unless explicitly stated.

\subsection{Repeated up-flavor quarks ($i=k=u,c$)}

\begin{align}
\hat U_{(12)}^{XXZ}(ijil)=\left(
\begin{array}{cc}
 1.98 & 0.01 \\
 -2.87 & 0.44 \\
\end{array}
\right),\ \ \ 
\hat U_{(13)}^{(LR/RL)Z}(ijil) =\left(
\begin{array}{cc}
 3.03 & 0. \\
 -1.44 & 0.87 \\
\end{array}
\right),
\end{align}
\begin{align}
\hat U_{(3)}^{XXZ}(iji\ell)=0.76,\ \ \
\hat U_{(44'55')}^{XXZ}(ijil) =
\left(
\begin{array}{cccc}
 0.85 & -0.17 &  -0.12 i &  -0.22 i \\
 -0.17 & 0.85 &  -0.22 i &  -0.12 i \\
 0.08 i & -0.10 i & 1.77 & -0.36 \\
  -0.10 i & 0.08 i & -0.36 & 1.77 \\
\end{array}
\right),
\end{align}
\begin{align}
\hat U_{(45)}^{(LR/RL)Z}(ijil)=
\left(
\begin{array}{cc}
 0.67 &  0.35 i \\
 0.02 i & 1.41 \\
\end{array}
\right),\ \ \ 
\hat U_{(6)}^{XXX}(ijil)=0.91,
\end{align}
\begin{align}
\hat U_{(6)}^{XXY}(ijil)=1.20,\ \ \ 
\hat U_{(67)}^{LRR/RLL}(ijil)=
\left(
\begin{array}{cc}
 0.95 &  0.08i  \\
 0  & 1.45 \\
\end{array}
\right),
\end{align}
\begin{align}
\hat U_{(77'8)}^{XXX}(ijil)=
\left(
\begin{array}{ccc}
 1.05  & -0.05 &-0.06 i \\
 -0.05  & 1.05 & 0.06 i \\
 0 & 0 & 0.69
\end{array}
\right).
\end{align}

\subsection{Repeated down-flavor quarks ($j=l=d,s,b$)}

\begin{align}
\hat U_{(12)}^{XXZ}(ijkj)=
\left(
\begin{array}{cc}
 1.98 & 0.01 \\
 -2.87 & 0.44 \\
\end{array}
\right),\ \ \ 
\hat U_{(13)}^{(LR/RL)Z}(ijkj)
= \left(
\begin{array}{cc}
 3.03 & 0. \\
 -1.44 & 0.87 \\
\end{array}
\right),
\end{align}
\begin{align}
\hat U_{(3)}^{XXZ}(ijkj)=0.76,\ \ \ 
\hat U_{(45)}^{XXZ}(ijkj)=\hat U_{(\overline 4\overline 5)}^{XXZ}(ijkj)=
\left(
\begin{array}{cc}
 0.67 &  -0.35 i \\
  -0.02 i & 1.41 \\
\end{array}
\right),
\end{align}
\begin{align}
\hat U_{(4\overline{4}5\overline{5})}^{(LR/RL)Z}(ijkj)=
\left(
\begin{array}{cccc}
 0.85 & -0.17 &  0.12 i &  0.22 i \\
 -0.17 & 0.85 &  0.22 i &  0.12 i \\
 -0.08 i & 0.10 i & 1.77 & -0.36 \\
  0.10 i & -0.08 i & -0.36 & 1.77 \\
\end{array}
\right),
\end{align}
\begin{align}
\hat U_{(6)}^{XXX}(ijkj)=1.20,\ \ \ 
\hat U_{(6)}^{LLR/RRL}(ijkj)=0.91,
\end{align}
\begin{align}
\hat U_{(6\overline{7})}^{LRR/RLL}(ijkj)=
\left(
\begin{array}{cc}
 0.95\, &  0.08 i \\
  0 &  1.45 \\
 \end{array}
\right),\ \ \
\hat U_{(7\overline{7}8)}^{XXX}(ijkj)=
\left(
\begin{array}{ccc}
 1.45  & -0.45 & -0.03 i \\
 -0.45  & 1.45 & 0.03 i \\
 0 & 0 & 0.69
\end{array}
\right).
\end{align}

\subsection{Repeated up- and down-flavor quarks ($i=k=u,c$ and $j=l=d,s,b$)}

\begin{align}
\hat U_{(12)}^{XXZ}(ijij)=\left(
\begin{array}{cc}
 1.98 & 0.01 \\
 -2.87 & 0.44 \\
\end{array}
\right),\ \ \ 
\hat U_{(13)}^{(LR/RL)Z}(ijij) =\left(
\begin{array}{cc}
 3.03 & 0. \\
 -1.44 & 0.87 \\
\end{array}
\right),
\end{align}
\begin{align}
\hat U_{(3)}^{XXZ}(ijij)=0.76,\ \ \ 
\hat U_{(45)}^{XXZ}(ijij)
=\left(
\begin{array}{cc}
 0.61  & -0.66 i \\
 -0.03 i & 1.26 \\
\end{array}
\right),
\end{align}
\begin{align}
\hat U_{(45)}^{(LR/RL)Z}(ijij)=\left(
\begin{array}{cc}
 0.73 & -0.04 i \\
 0.01 i & 1.57 \\
\end{array}
\right),\ \ \ 
\hat U_{(67)}^{LRR/RLL}(ijij)=\left(
\begin{array}{cc}
 0.95 & 0.08 i \\
 0  & 1.45 \\
\end{array}
\right).
\end{align}

\subsection{All different flavor quarks ($i\neq k$, $j \neq l$)}

\begin{align}
\hat U_{(11'22')}^{XXZ}(ijkl)=
\left(
\begin{array}{cccc}
 3.26 & -1.28 & -0.06 & 0.07 \\
 -1.28 & 3.26 & 0.07 & -0.06 \\
 -0.78 & -2.10 & 0.73 & -0.29 \\
 -2.10 & -0.78 & -0.29 & 0.73 \\
\end{array}
\right),
\end{align}
\begin{align}
\hat U_{(13')}^{(LR/RL)Z}(ijkl)
=\left(
\begin{array}{cc}
 3.03 & 0 \\
 -1.44 & 0.87\\
\end{array}
\right),\ \ \ 
\hat U_{(33')}^{XXZ}(ijkl)=\left(
\begin{array}{cc}
 1.25 & -0.49 \\
 -0.49 & 1.25 \\
\end{array}
\right),
\end{align}
\begin{align}
\hat U_{(44'55')}^{XXZ}(ijkl) =
=
\left(
\begin{array}{cccc}
 0.85 & -0.17 &  -0.12 i &  -0.22 i \\
 -0.17 & 0.85 &  -0.22 i &  -0.12 i \\
 0.08 i & -0.10 i & 1.77 & -0.36 \\
  -0.10 i & 0.08 i & -0.36 & 1.77 \\
\end{array}
\right), 
\end{align}
\begin{align}
\hat U_{(4\overline{4}'5\overline{5}')}^{(LR/RL)Z}(ijkl) 
=\left(
\begin{array}{cccc}
 0.85 & -0.17 &  0.12 i &  0.22 i \\
 -0.17 & 0.85 &  0.22 i &  0.12 i \\
 -0.08 i & 0.10 i & 1.77 & -0.36 \\
  0.10 i & -0.08 i & -0.36 & 1.77 \\
\end{array}
\right),
\end{align}
\begin{align}
\hat U_{(6{6'})}^{XXX}(ijkl)=
\left(
\begin{array}{cc}
 1.08 &  0.19  \\
 0.19  & 1.08 \\
\end{array}
\right),\ \ \ 
\hat U_{(6{6'})}^{XXZ}(ijkl)=
\left(
\begin{array}{cc}
 1.08 &  -0.19  \\
 -0.19  & 1.08 \\
\end{array}
\right),
\end{align}
\begin{align}
\hat U_{(6\overline{7'})}^{LRR/RLL}(ijkl)=
\left(
\begin{array}{cc}
 0.95 &  0.08 i \\
  0 &  1.45 \\
 \end{array}
\right),
\end{align}
\begin{align}
\hat U_{(7\overline{7}\overline{7}'88')}^{XXX}(ijkl)=
\left(
\begin{array}{cccccc}
 1.44 & 0.02 & -0.07& -0.39 & -0.05 i & 0.01 i\\
 0.02 & 1.44 & -0.39 & -0.07 & 0.01 i & -0.05 i\\
 -0.07 & -0.39 & 1.44 & 0.02 & 0.05 i & -0.01 i\\
 -0.39 &  -0.07 & 0.02 & 1.44 & -0.01 i & 0.05 i \\
 0 &  0 & 0 & 0 & 0.69 & 0\\
 0 & 0 & 0 & 0 & 0 & 0.69\\
\end{array}
\right).
\end{align}

\section{Numerical Constants}\label{app:constants}

For the numerical calculations, we use the constants shown in Tables\,(\ref{tab:constantsmasses}) and (\ref{tab:constantslifetimeanddc}).

\begin{table}
\centering
\begin{tabular}{c|c||c|c||c|c}
    Constant & Value & Constant & Value & Constant & Value\\
\hline\hline
$m_u$ & 2.16\,MeV & $m_\pi$ & 139.57039\,MeV & $m_e$ & 0.51099895\,MeV\\
$m_d$ & 4.67\,MeV & $m_K$ & 493.677\,MeV & $m_\mu$ & 105.6583755\,MeV\\
$m_s$ & 93.4\,MeV & $m_D$ & 1.86966\,GeV & $m_\tau$ & 1776.86\,MeV\\
$m_c$ & 1.27\,GeV & $m_{D_s}$ & 1.96835\,GeV  & & \\
$m_b$ & 4.18\,GeV & $m_B$ & 5.27934\,GeV & &\\
\hline
\hline
\end{tabular}
\caption{Values of the masses used in our calculations. Taken from Ref.\,\cite{Workman:2022ynf}.}\label{tab:constantsmasses}
\end{table}

\begin{table}
\centering
\begin{tabular}{c|c||c|c}
    Constant & Value & Constant & Value\\
\hline
\hline
$\tau_{K}$ & $1.238\times 10^{-8}\,\mbox{s}$ & $f_\pi$ &  130.2\,MeV\\
$\tau_{D}$ & $1.033\times 10^{-12}\,\mbox{s}$ & $f_K$ & 155.7\,MeV\\
$\tau_{D_s}$ & $5.04\times 10^{-13}\,\mbox{s}$ & $f_D$ & 212\,MeV\\
$\tau_{B}$ & $2.903\times 10^{-13}\,\mbox{s}$ & $f_B$ & 190\,MeV\\
$\tau_{\tau}$ & $2.903\times 10^{-13}\,\mbox{s}$ & & \\
\hline
\hline
\end{tabular}
\caption{Lifetimes and meson decay constants used in our numerical calculations. Taken from Ref.\,\cite{Workman:2022ynf}.}\label{tab:constantslifetimeanddc}
\end{table}

\end{appendices}

\bibliographystyle{unsrt} 
\bibliography{bib}

\end{document}